\documentclass[useAMS,usenatbib]{mnras}
\usepackage{graphicx}
\usepackage{rotating}
\usepackage{amsmath}
\usepackage{longtable}
\usepackage{subfig}
\usepackage{float}
\usepackage{amssymb}
\usepackage[version=3]{mhchem}
\newif\ifAMStwofonts

\newcommand{\co}{C$^{18}$O}
\newcommand{\HH}{H$_2$}
\newcommand{\Rd}{R_{dep}}

\title[CO Depletion - G351.77-0.51]{On the size of the CO-depletion radius in the IRDC G351.77-0.51}
\author[G.~Sabatini et al.]
{\parbox{\textwidth}{\raggedright G. Sabatini$^{1, 2, 3}$\thanks{E-mail: sabatini@ira.inaf.it \textit{or} giovanni.sabatini4@unibo.it},
A.~Giannetti$^{2}$,
S.~Bovino$^{3}$,
J.~Brand$^{2}$,
S.~Leurini$^{4}$,
E.~Schisano$^{2, 5}$,
T.~Pillai$^{6, 7}$,
and K.~M.~Menten$^{6}$}
\vspace{0.4cm}\\
$^{1}$Dipartimento di Fisica e Astronomia, Universit\`{a} degli Studi di Bologna, via Gobetti 93/2, I-40129 Bologna, Italy\\
$^{2}$INAF - Istituto di Radioastronomia - Italian node of the ALMA Regional Centre (ARC), via Gobetti 101, I-40129 Bologna, Italy\\
$^{3}$Departamento de Astronom\'ia, Universidad de Concepci\'on, Barrio Universitario, Concepci\'on, Chile\\
$^{4}$INAF - Osservatorio Astronomico di Cagliari, via della Scienza 5, 09047 Selargius (CA), Italy\\
$^{5}$Istituto di Astrofisica e Planetologia Spaziali - INAF, Via Fosso del Cavaliere 100, I-00133 Roma, Italy\\
$^{6}$Max-Planck-Institut f\"ur Radioastronomie, Auf dem H\"ugel 69, 53121 Bonn, Germany\\
$^{7}$Institute for Astrophysical Research, 725 Commonwealth Ave, Boston University Boston, MA 02215, USA}
\begin{document}

\date{Accepted 2019 September 30. Received 2019 September 25; in original form 2019 March 22}

\pagerange{\pageref{firstpage}--\pageref{lastpage}} \pubyear{2017}

\maketitle

\label{firstpage}

\begin{abstract}
An estimate of the degree of CO-depletion ($f_D$) provides information on the physical conditions occurring in the innermost and densest regions of molecular clouds.
A key parameter in these studies is the size of the depletion radius, i.e. the radius within which the C-bearing species, and in particular CO, are largely frozen onto dust grains. A strong depletion state (i.e. $f_D>10$, as assumed in our models) is highly favoured in the innermost regions of dark clouds, where the temperature is $<20$~K and the number density of molecular hydrogen exceeds a few $\times$10$^{4}$~cm$^{-3}$. 
In this work, we estimate the size of the depleted region by studying the Infrared Dark Cloud (IRDC) G351.77-0.51. Continuum observations performed with the {\it Herschel Space Observatory} and the {\it LArge APEX BOlometer CAmera}, together with APEX \co\: and C$^{17}$O J=2$\rightarrow$1 line observations, allowed us to recover the large-scale beam- and line-of-sight-averaged depletion map of the cloud. We built a simple model to investigate the depletion in the inner regions of the clumps in the filament and the filament itself. The model suggests that the depletion radius ranges from 0.02 to 0.15 pc, comparable with the typical filament width (i.e.$\sim$0.1~pc). At these radii, the number density of \HH~reaches values between 0.2 and 5.5$\times$10$^{5}$~cm$^{-3}$. 
These results provide information on the approximate spatial scales on which different chemical processes operate in high-mass star forming regions and also suggest caution when using CO for kinematical studies in IRDCs.
\end{abstract}

\begin{keywords}
Astrochemistry $<$ Physical Data and Processes -- Molecular processes $<$ Physical Data and Processes -- ISM: Molecules $<$ Interstellar Medium (ISM), Nebulae -- Stars: formation $<$ Stars -- Galaxy: abundances $<$ The Galaxy -- submillimetre: ISM $<$ Resolved and unresolved sources as a function of wavelength
\end{keywords}

\section{Introduction}
\label{sec_intro}
\noindent
In the last two decades, evidence of CO depletion has been widely found in numerous sources (e.g. \citealt{Kramer99, Caselli99, Bergin02, Fontani12, Wiles16}), with important consequences for the chemistry, such as an increase in the fraction of deuterated molecules (\citealt{Bacmann03}). 
Cold and dense regions within molecular clouds with temperatures, $T$, $< 20$~K (e.g. \citealt{Caselli08}) and number densities, $n$(\HH), $>$ a few $\times$ 10$^{4}$ cm$^{-3}$ (e.g. \citealt{Crapsi05}; see also \citealt{BerginTafalla07} and references therein), provide the ideal conditions that favour depletion of heavy elements onto the surface of dust grains. Here, chemical species can contribute not only in the gas-phase chemistry, but also in surface reactions by being frozen onto the dust grains. This chemical dichotomy can lead to the formation of complex molecules, modifying the chemical composition over a significant volume of a cloud (e.g. \citealt{Herbst&vanDishoeck09}).\\
\indent How much of the CO is depleted onto the surface of dust grains is usually characterized by the depletion factor (e.g. \citealt{Caselli99,Fontani12}), defined as the ratio between the ``expected'' CO abundance with respect to \HH~($\chi^E_{\rm CO}$) and the observed one ($\chi^O_{\rm CO}$):

\begin{equation}\label{eq:depletionfactor}
f_D=\frac{\chi^E_{\rm CO}}{\chi^O_{\rm CO}}.
\end{equation}

\noindent
The main CO isotopologue (i.e. \ce{^{12}C^{16}O}) is virtually always optically thick, and thus its intensity is not proportional to the CO column density.
One of the less abundant CO isotopologues (i.e. \co, as in this work) can be used to obtain a much more accurate estimate of $f_D$.\\
\indent In general, in dense and cold environments, such as massive clumps in an early stage of evolution, the estimate of the depletion degree provides important information such as the increase of the efficiency of chemical reactions that occur on the dust grain surface, favoured by the high concentration of the depleted chemical species.\\
High-mass star forming regions are potentially more affected by large-scale depletion: the high volume densities of \HH~both in the clumps and in the surrounding intra-clump regions make these sources more prone to high levels of molecular depletion (e.g. \citealt{Giannetti14}) because the timescale after which depletion becomes important (see, e.g. \citealt{Caselli99,Caselli02}) decreases with increasing volume density. In fact, the depletion time scale ($\tau_{dep}$) is inversely proportional to the absorption rate, $\kappa_{abs} = \sigma\langle v \rangle n_g S$ [s$^{-1}$], of the freezing-out species: where $\sigma$ is the dust grains cross-section, $\langle v \rangle$ is the mean velocity of the Maxwellian distribution of gaseous particles, $n_g$ is the dust grains number density, and S is the sticking coefficient (see also \citealt{BerginTafalla07} for more details).\\
\indent In different samples of young high-mass star-forming regions, the observed depletion factors vary between $\approx1$, in the case of complete absence of depletion, and a few tens (e.g. in \citealt{Thomas08}, \citealt{Fontani12} and \citealt{Feng16}). The largest values reported are those in \cite{Fontani12} where $f_D$ can reach values of up to 50-80, a factor $\sim 10$ larger than those observed in samples of low mass clouds (e.g. \citealt{Bacmann02, Bacmann03, Ceccarelli07, Christie12}). \cite{Giannetti14} noted that the method used by \cite{Fontani12} to calculate N(\HH) yielded values $\sim˘2.7$ times larger than that they themselves used to obtain this quantity, mainly due to the different dust absorption coefficient, $\kappa$, assumed (i.e. an absorption coefficient at 870 $\mu$m, $\kappa_{870 \mu\rm{m}} = 1.8$ cm$^2$ g$^{-1}$ assumed by \citealt{Giannetti14}, while $\kappa_{870 \mu\rm{m}}\sim 0.77$ cm$^2$ g$^{-1}$ assumed by \citealt{Fontani12} and derived here following \citealt{Beuther05} and \citealt{Hildebrand83}).\\
\indent Rescaling the depletion factors of these sources under the same assumptions, we note that the typical values ​​of $f_D$ vary from a minimum of 1 to a maximum of $\sim$10, except for some particular cases. However, there are studies, which were made with much higher resolution, that reported even larger depletion factors: this is the case for the high-mass core in the IRDC G28.34+0.06 where, at a spatial resolution of $\sim 10^{-2}$ pc (using a source distance of 4.2 kpc; \citealt{Urquhart18}), $f_D$ reaches values of $10^2 - 10^3$, the highest values of $f_D$ found to date (\citealt{Zhang09}).\\
\indent Observing depletion factors ranging from 1 to 10, means that along the line of sight there will be regions in which the CO is completely in gas phase (i.e. not depleted), and other regions in which the depletion factor reaches values larger than 10. Knowledge of the size of the region within which most of the CO is largely frozen onto dust grains (the depletion radius) provides information on the approximate spatial scales on which different chemical processes operate in high-mass star forming regions. At these scales, for example, it is reasonable to think that derivation of \HH~column densities from CO lines (see e.g. \citealt{Bolatto13} and reference therein) and the studies of the gas-dynamics using the carbon monoxide isotopologues lines are strongly affected by depletion process, as not all of the CO present is directly observable. 
Furthermore, the disappearance of CO from the gas phase also favours the deuteration process (e.g. \citealt{Roberts00, Bacmann03}). This is due to the following reaction which becomes increasingly inefficient, while the amount of frozen CO increases:

\begin{equation}\label{reaction1}
  {\rm H^+_3 + CO \rightarrow HCO^+ + H_2} .
\end{equation}

\noindent 
In a high CO-depletion regime, ${\rm H_3^+}$ is destroyed less efficiently by reaction~(\ref{reaction1}), remaining available for other reactions. Deuterium enrichment in the gas phase is enhanced by the the formation ${\rm H_2D^+}$ through the reaction: %dissociative recombination of ${\rm H_2D^+}$ (i.e. $H_2D^+ + e^- \rightarrow HD + H$ or $H_2 + D$; see \citealt{Aikawa13}), which is formed through the reaction:
 
\begin{equation}\label{reaction2}
  {\rm H_3^+ + HD \rightleftharpoons H_2D^+ + H_2 + \Delta E}\,,
\end{equation}

\noindent
which proceeds from left-to-right, unless there is a substantial fraction of ortho-\HH~(e.g. \citealt{Gerlich02}), the only case in which reaction (\ref{reaction2}) is endothermic and naturally proceeds in the direction in which the observed deuterated fraction decreases. For this reason, for a higher degree of CO depletion, the formation of ${\rm H_2D^+}$ by reaction~(\ref{reaction2}) is more efficient due to the greater amount of ${\rm H_3^+}$ available.\\
\indent Single cell and one-dimensional numerical calculations (e.g. \citealt{Sipila13, Kong15}) are fast enough to include detailed treatment of the depletion process together with large comprehensive networks to follow the chemistry evolution in these environments.\\
\indent However, when we move to more accurate three-dimensional simulations, strong assumptions are needed to limit the computational time and allow the study of the dynamical effects on a smaller set of chemical reactions. \cite{Koertgen17, Koertgen18} for example, analysed the dependence of the deuteration process on the dynamical evolution, exploring how the chemical initial conditions influence the process. They performed 3D magnetohydrodynamical (MHD) simulations fully coupled for the first time with a chemical network that describes deuterium chemistry under the assumption of complete depletion (i.e. $f_D = \infty$) onto a spatial scale of $\sim 0.1$ pc from the collapse centre. However, the real extent of the depletion radius is still unknown, and this uncertainty might alter the theoretical models that describe the chemical evolution of star-forming regions.\\
\indent One way to shed light on these issues, is by mapping high-mass star-forming regions on both large and small scales with the goal to determine the depletion factor fluctuations over a broad range of densities and temperatures.\\
\indent In this paper, we selected the nearest and most massive filament found in the $870 \mu$m {\it APEX Telescope Large AREA Survey of the Galaxy} (ATLASGAL) (see \citealt{Schuller09} and \citealt{Li16}), the IRDC G351.77-0.51, presented in Sect.~\ref{sec:datapresentation}. In Sect.~\ref{sec:results}, we present the dust temperature map obtained by fitting the Spectral Energy Distribution (SED) using continuum observations of the {\it Herschel} Infrared Galactic Plane Survey (Hi-GAL, \citealt{Molinari10}). In addition, we compute the column density maps of \HH\:and \co. Combining the \HH\:and \co\:column density maps allows us to produce the large-scale depletion map of the entire source. We evaluate variations of depletion over the whole structure with the aim to understand which scales are affected by depletion the most. By assuming a volume density distribution profile for the \HH\: and a step-function to describe the abundance profile of \co\:with respect to \HH, we estimate the size of the depletion radius (Sect.~\ref{discussion}). In Sect.~\ref{caveats} we finally explore how much the estimates of the depletion radius are affected by the model's assumptions.

\begin{figure}
\centering
\includegraphics[width=1\columnwidth]{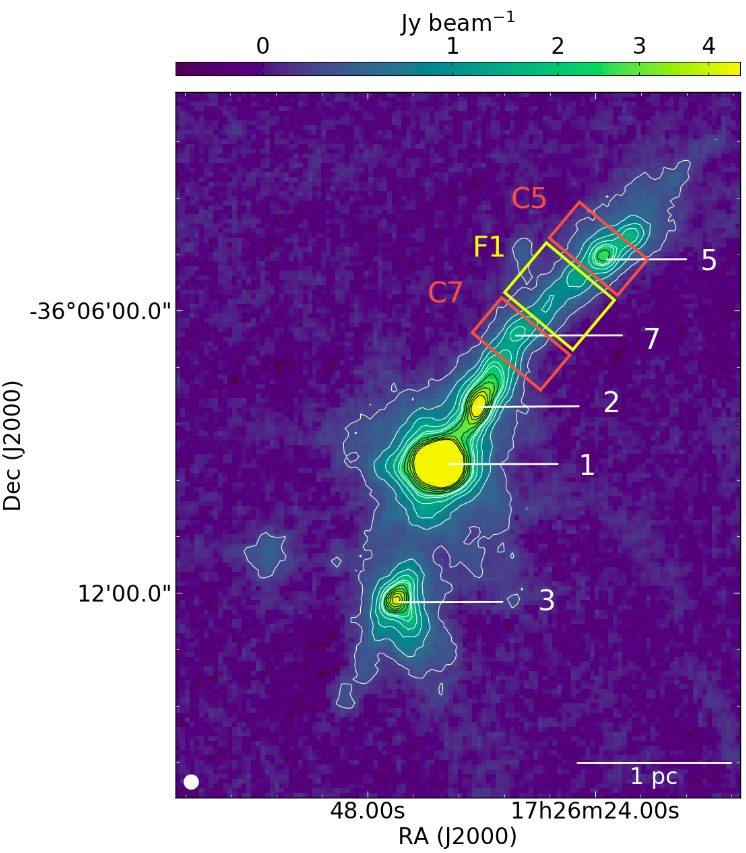}
\caption{{\it LArge APEX BOlometer CAmera} (LABOCA) map of the 870$\mu$m dust continuum emission from the IRDC G351.77-0.51. The white contours are from 5$\sigma$ (0.2 Jy beam$^{-1}$) to 2 Jy beam$^{-1}$ while the black contours are from 2.4 Jy beam$^{-1}$ to 4 Jy beam$^{-1}$ in steps of 10$\sigma$. The white labels indicate some of the dust clumps identified by \citet{Leurini11}. Orange and yellow boxes (i.e., regions C5, C7 and F1) are linked to the models discussed in Sect.~\ref{discussion}.}
\label{fig:leurini}
\end{figure}

\section{SOURCE AND DATA: IRDC G351.77-0.51}\label{sec:datapresentation}
\noindent
In Fig.~\ref{fig:leurini} we show a continuum image at 870 $\mu$m of G351.77-0.51 from the ATLASGAL survey (\citealt{Li16}), in which the clumps (labels 1, 2, 3, 5 and 7) appear well-pronounced along the filamentary structure of the star-forming region. Among the twelve original clumps defined in \citet{Leurini11}, we selected only the five clumps for which data of both C$^{17}$O and \co~are available (see Sect. \ref{opacity}).\\
\indent G351.77-0.51 is the most massive filament in the ATLASGAL survey within 1 kpc from us. 
\citealt{Leurini19} estimate M$\sim$ 2600 M$_\odot$ (twice that listed by \citealt{Li16} who used different values for dust temperature and opacity). 
Following the evolutionary sequence of massive clumps defined by \cite{Konig17} - originally outlined by \cite{Giannetti14, Csengeri16} - the main clump of G351.77-0.51 was classified as an infrared-bright (IRB) source\footnote{This class of object shows a flux larger than 2.6 Jy at 21-24 $\mu$m and no compact radio emission at 4-8 GHz within 10'' of the ATLASGAL peak.}, and revealed hot-cores features (see \cite{Giannetti17_june}. 
\cite{Leurini11_bis, Leurini19} studied the velocity field of the molecular gas component. They discussed the velocity gradients by considering whether they might be due to rotation, or outflow(s) around clump-1, or indicative of multiple velocity components detected in several \co\:spectra.\\
\indent In this paper we assume the same nomenclature as \cite{Leurini19} to identify the structures that compose the complex network of filaments of G351.77-0.51: below, we will refer to the central filament as ``main body'' or ``main ridge''. It is well-visible in Fig.~\ref{fig:leurini} as the elongated structure that harbours the five clumps, identified by white labels. The gas that constitutes the main body is cold and chemically young, in which \cite{Giannetti19} find high abundances of o-H$_2$D$^+$ and N$_2$D$^+$ and an age $\lesssim 10^5$ yr for the clump-7 (see that paper for more details). The northern part of the main body appears in absorption against the mid-IR background of our Galaxy up to 70 $\mu$m and in emission in the (sub)millimetre range (\citealt{Faundez04}). The analysis of the J$=$2$\rightarrow$1 \co~molecular line (see Sect.~\ref{C18O}) confirmed the presence of less dense molecular sub-filaments, linked to the main body: we refer to them as ``branches'', following the naming convention of \cite{Schisano14}.

\begin{figure*}
\centering
%\subfloat[][\emph{}]
{\includegraphics[width=.44\textwidth]{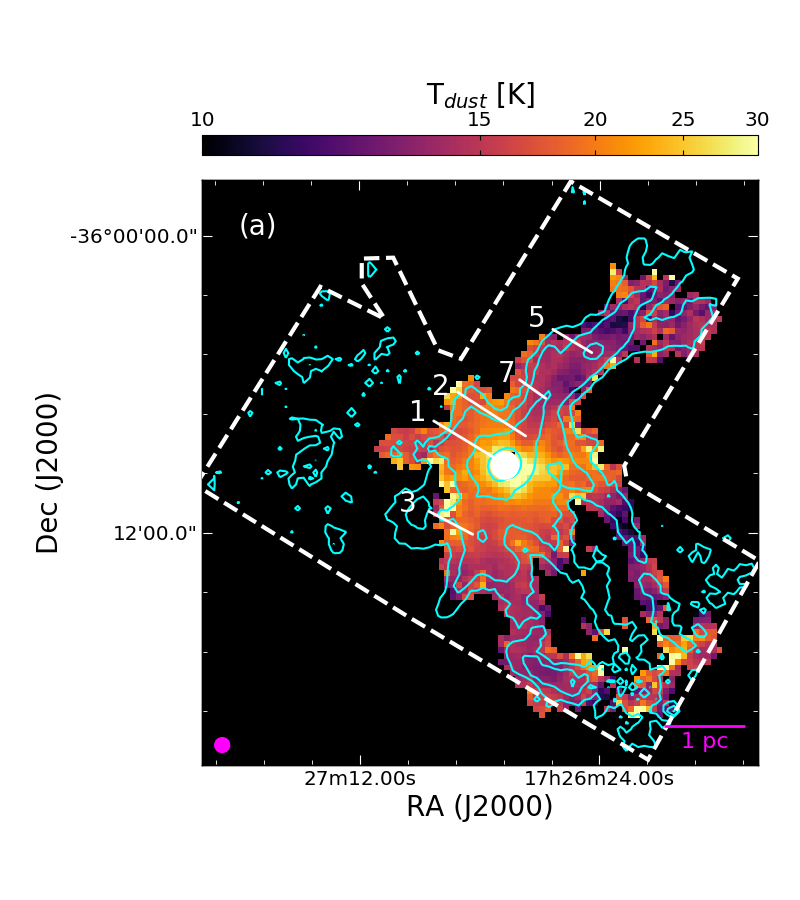}}
%\subfloat[][\emph{\co J(2-1) excitation temperature map}]
%{\includegraphics[width=.32\textwidth]{Tex.png}}
%\subfloat[][\emph{}]
{\includegraphics[width=.44\textwidth]{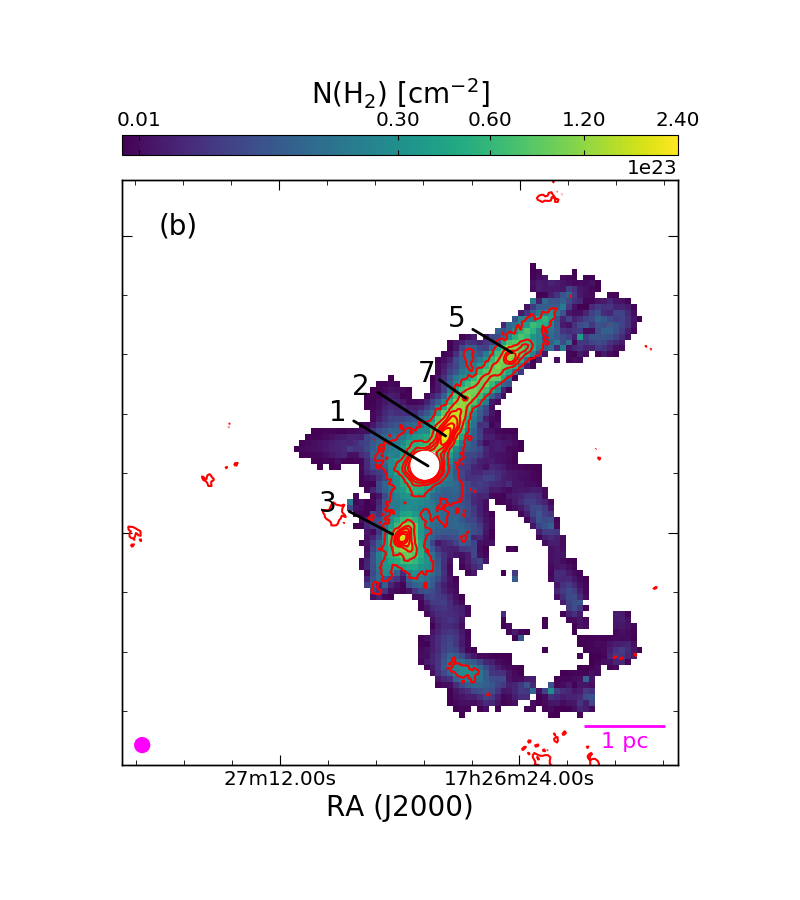}} 
\caption{(a) Dust temperature map from \citet{Leurini19}, generated by a pixel-by-pixel SED-fitting of the 160-500 $\mu$m continuum fluxes of the Hi-Gal Survey. The white dashed lines mark the region observed in \co~with APEX (Sect.~\ref{c18o_observations}). Cyan contours are defined on the integrated intensity map of the APEX \co\:J$=2\rightarrow 1$ line in Fig.~\ref{fig:OPACITY} (a) at 3, 9, 27 and 81$\sigma$ contours, where 3$\sigma =$ 0.9 K km s$^{-1}$; (b) \HH\:column density map from the pixel-by-pixel SED-fitting from \citet{Leurini19}, scaled to a gas mass-to-dust ratio equal to 120 (see text). Red contours are the same of the LABOCA map of the 870$\mu$m dust continuum emission shown in Fig.~\ref{fig:leurini}. In both panels, we masked the saturated hot-core region in {\it Herschel} data (white circle).}
\label{fig:Tmaps}
\end{figure*}

\subsection{\co\:map}\label{c18o_observations}
The \co\:J$=$2$\rightarrow$1 observations were carried out with the {\it Atacama Pathfinder Experiment} 12-meter submillimeter telescope (APEX) between 2014 August and November. The observations were centred at 218.5 GHz, with a velocity resolution of 0.1 km~s$^{-1}$. The whole map covers an approximate total area of 234 ${\rm arcmin}^2$. We refer to \cite{Leurini19}, for a more detailed description of the dataset.\\
The root mean square (rms) noise was calculated iteratively in each pixel because it is not uniform in the map, due to the different exposure times. The first estimate of rms noise was obtained from the unmasked spectra, then, any emission higher than 3$\sigma$ was masked and the rms was recomputed. The cycle continues until the difference between the rms noise of two consecutive iterations (i.e. $\sigma_{i+1} -\sigma_{i}$) is equal to 10$^{-4}$ K. We estimate a typical-final noise of 0.30 K (between 0.15 and 0.45 K) per velocity channel.

\section{Results}\label{sec:results}
\subsection{Temperature and \HH~column density maps}\label{temperature_maps}
Calculating \co\:column densities requires the gas excitation temperature, $T_{\rm ex}^{\rm C^{18}O}$. We derived $T_{\rm ex}^{\rm C^{18}O}$ from the dust temperature, $T_{\rm dust}$, map, which can be obtained from the {\it Herschel} data through pixel-by-pixel SED fits. This allows one to determine also the \HH\:column density map from the emission between 160 and 500 $\mu$m.\\
\indent In this work we adopt the dust temperature map presented in \cite{Leurini19}. To estimate the dust temperature of the whole filamentary structure of G351.77-0.51, these authors used the {\it Herschel} (\citealt{Pilbratt10}) Infrared Galactic Plane Survey (Hi-GAL, \citealt{Molinari10}) images at 500, 350 and 250 $\mu$m from SPIRE (\citealt{Griffin10}) and at 160 $\mu$m from PACS (\citealt{Poglitsch10}). 
The authors adopted a model of two emission components (Schisano et al. $subm.$), splitting the fluxes observed by {\it Herschel} in each pixel into the filament and background contributions, as discussed by \cite{Peretto10}. Then, they fitted the filament emission pixel-by-pixel with a grey body model deriving the dust temperature and the \HH~column density. 
They assumed a dust opacity law $\kappa_0(\nu/\nu_0)^{\beta}$ with $\beta=2$, $\kappa_0=0.1$ cm$^2$ g$^{-1}$ at $\nu_0=1250$ GHz (\citealt{Hildebrand83}). This prescription assumes a gas-to-dust ratio equal to 100.
A detailed description of the procedure is given in \cite{Leurini19}.\\
\indent Their maps, shown in Fig.~\ref{fig:Tmaps} have a resolution of $36''$, i.e. the coarsest resolution in {\it Herschel} bands, that is comparable to the resolution of our \co\:data ($29''$). In most of the map, the dust temperature (panel (a) of Fig.~\ref{fig:Tmaps}) ranges between 10 and 30 K, while along the main body and in the region to the south of clump-3 in Fig.~\ref{fig:leurini} typically $T_{dust}\lesssim$ 15 K are found.\\
\indent $T_{\rm ex}^{\rm C^{18}O}$ was derived from $T_{\rm dust}$ by applying the empirical relation defined by \cite{Giannetti17_june}, who suggest a relation $T_{\rm ex}^{\rm C^{18}O}=$ 1.46$\times T_{\rm dust} -12.9$ with an intrinsic scatter of 6.7 K. However, this relation is only valid for dust temperatures between $\sim 10$ and $\sim 45$ K, while at low $T_{\rm dust}$ it may underestimate $T_{\rm ex}^{\rm C^{18}O}$. Our data are concentrated in this low temperature regime, where $T_{\rm dust}\lesssim$ 9 K would be translated into negative $T_{\rm ex}^{\rm C^{18}O}$ following the relation mentioned earlier. Therefore we decided to use the flattest curve allowed by the uncertainties at 2$\sigma$ of the original fit to limit this issue: $T_{\rm ex}^{\rm C^{18}O}=$ $T_{\rm dust} -2.6$ and imposing a lower limit of $T_{\rm ex}=$ 10.8 K, equal to the separation between the levels of the J$=$2$\rightarrow$1 transition of the \co~line.\\
$T_{\rm ex}^{\rm C^{18}O}$ typically ranges between $\sim$ 11 and 28 K with peaks of up to 40 K. Most of the pixels along the main ridge show a temperature of 12-13 K. At these temperatures CO is efficiently removed from the gas phase and frozen onto dust grains for densities exceeding few $\times$ 10$^4$~cm$^{-3}$ (see \citealt{Giannetti14}). This suggests that the depletion factor in these regions will have the highest values of the entire IRDC (as confirmed by the analysis of the depletion map - Sect.~\ref{observeddepletion}).\\

\begin{figure*}
\centering
{\includegraphics[width=1\textwidth]{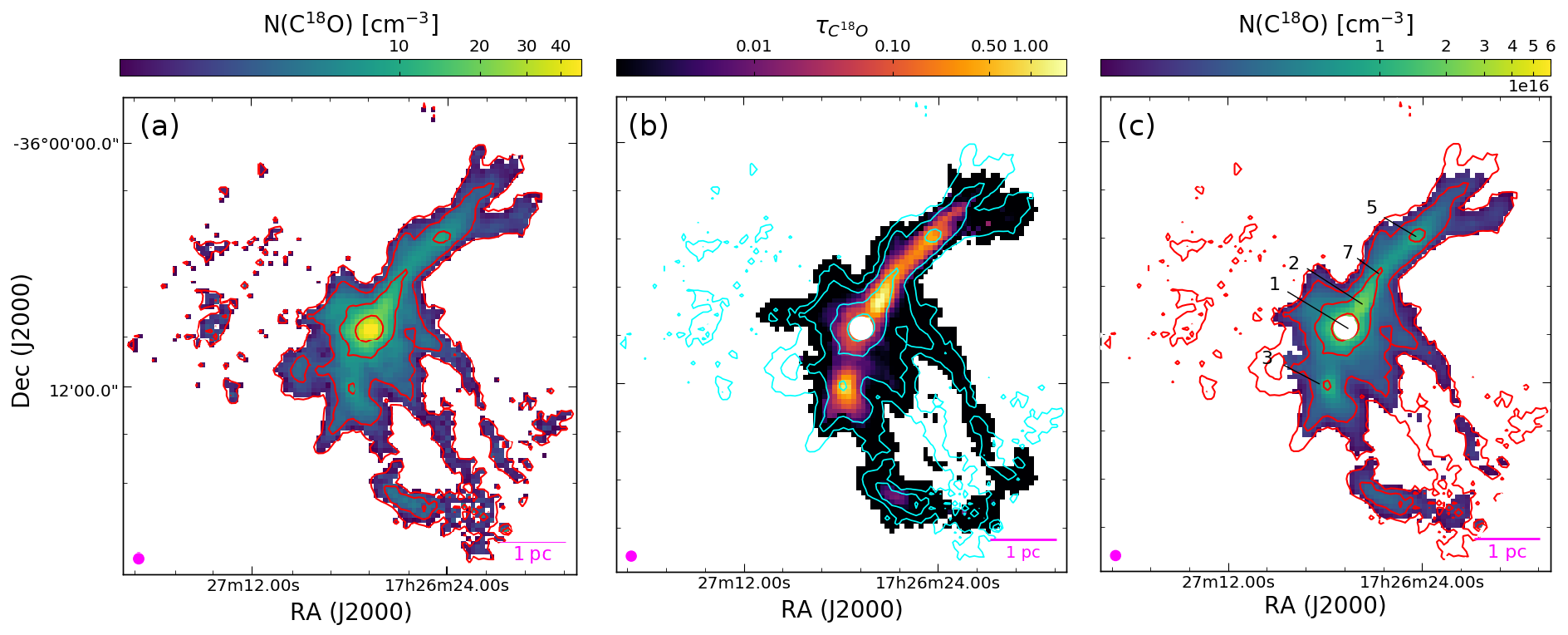}} 
\caption{(a) Integrated intensity distribution of the \co\:J=2$\rightarrow$1 molecular line transition detected with {\it APEX} in units of K km s$^{-1}$. (b) Map of the opacity from the N(\HH) map, following the procedure described in Sect.~\ref{opacity}. The final $\tau$-N(\HH) relation used is: log$_{10}$($\tau_{\rm C^{18}O}$)$=$2.5$~$log$_{10}$(N(\HH))-57.4. (c) \co\:column density map obtained from the integrated intensity distribution, shown in Fig.~\ref{fig:OPACITY} (a), using eq.~(\ref{eq:nc18o}). The red and cyan contours are the 3, 9, 27 and 81$\sigma$ contours of the integrated flux density distribution in panel (a), where 3$\sigma =$ 0.9 K km s$^{-1}$.}
\label{fig:OPACITY}
\end{figure*}

\indent We modified the \HH~column density map of \cite{Leurini19} by using a different value of gas mass-to-dust ratio, $\gamma$, equal to 120 as derived by \cite{Giannetti17_oct}, by assuming a galactocentric distance $D_{GL} = 7.4$ kpc (\citealt{Leurini19}). The resulting \HH~column density map is shown in Fig.~\ref{fig:Tmaps} ($b$). The saturated region in {\it Herschel} data, corresponding to the hot-core clump, was masked out and excluded from our analysis. In the coldest regions of Fig. \ref{fig:Tmaps} (a), the molecular hydrogen reaches a column density of 2.4 $\times$ 10$^{23}$ cm$^{-2}$, decreasing to 4.2 $\times$ 10$^{21}$ cm$^{-2}$ in the regions in which the dust is warmer. Fig. \ref{fig:Tmaps} (b) shows that the high-column density material (i.e. $\gtrsim 1 \times 10^{23}$ cm$^{-2}$) is distributed in a single filamentary feature and in clump-like structures, the so called ``main-body'' in \cite{Leurini19}.
%, changed by several over-dense areas where star-forming events are favoured.

\subsection{Column densities of \co~}\label{C18O}
Under the assumption of local thermal equilibrium (LTE), we derived the \co\:column density, N(\co), from the integrated line intensity of the (2-1) transition. Following \cite{Kramer91}, the general formula for ${\rm N}({\rm C^{18}O})$ is:

\begin{multline}\label{eq:nc18o}
{\rm N}({\rm C^{18}O}) = \frac{C_\tau}{\eta_c} \frac{3h}{8\pi^3\mu^2} \frac{Z}{2} e^{\frac{E_{low}}{k_B T_{ex}}} \left[ 1- e^{-\frac{h\nu_{J,J-1}}{k_B T_{ex}}} \right]^{-1} \\
\times [J(T_{ex}, \nu_{\rm C^{18}O}) - J(T_{bg}, \nu_{\rm C^{18}O})]^{-1} \int T_{MB}\: d\upsilon \\
= C_\tau\:f(T_{ex}) \int T_{MB}\: d\upsilon,\\
\end{multline}

\noindent
where $C_\tau$ is the optical depth correction factor defined as $\tau/[1 - {\rm exp}(-\tau)]$ and the $\tau$ is the optical depth of the J$=$2$\rightarrow$1 transition of \co (see Sect.~\ref{opacity}); $h$ and $k_B$ are the Planck and Boltzmann constants, respectively; $\eta_c$ is the {\it filling factor}; $\mu$ is the dipole moment of the molecule; $Z$ is the partition function; $E_{low}$ is the energy of the lower level of the transition and $\nu_{J,J-1}$ is the frequency of the $J\rightarrow J-1$ transition of the considered molecule (in this case, the \co~J$=$2$\rightarrow$1, equal to $\sim$219.5 GHz); $J(T_{\rm ex},\nu)= (h\nu/k_B)($exp$(h\nu/k_B T)-1)^{-1}$; $T_{bg}\cong2.7$ K, is the background temperature; $T_{MB}$ is the main beam temperature, and its integral over the velocity range covered by the \co\:line is shown in Fig.~\ref{fig:OPACITY} ($a$). In the last line of eq.~(\ref{eq:nc18o}), $f({T}_{ex}$) incorporates all the constants and the terms depending on $T_{ex}$. We further considered possible saturation effects of the continuum {\it Herschel} maps, in the hot-core region (i.e. clump-1 in Fig.~\ref{fig:leurini}).

In the following paragraphs, we discuss the steps that allowed us to derive the $C_\tau$ map (Fig.~\ref{fig:OPACITY}$b$) and the final \co\:column density map (Fig.~\ref{fig:OPACITY}$c$), necessary to produce the depletion map discuss in Sect.~\ref{observeddepletion}.

\subsubsection{Opacity correction}\label{opacity}

\begin{figure}
\centering
\includegraphics[width=1\columnwidth]{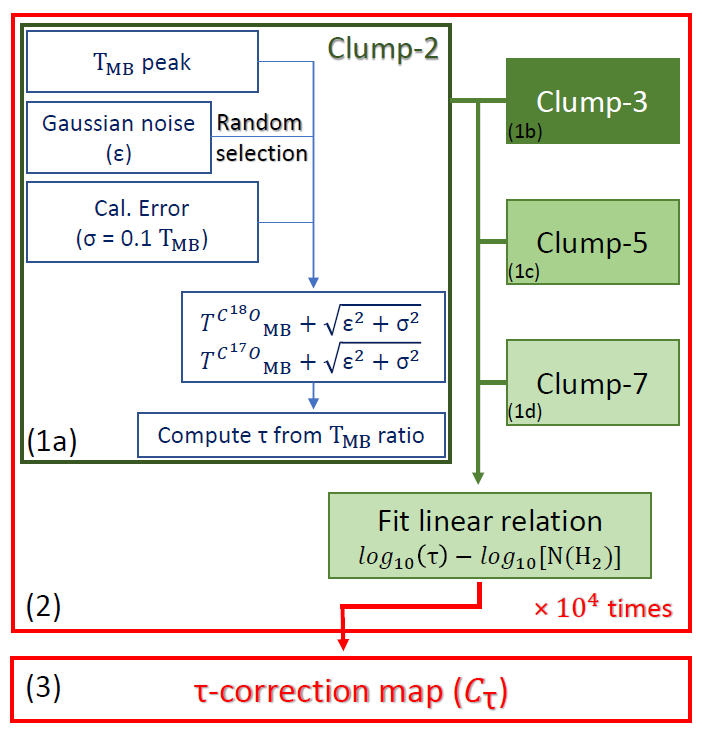}
\caption{Schematic diagram showing how the correction $C_\tau$ was calculated starting from the {\rm \co} and {\rm C$^{17}$O} single pointing observation of clumps 2, 3, 5 and 7 in \citet{Leurini11}.}
\label{fig:scheme}
\end{figure}

We estimate the optical depth of \co~J$=$2$\rightarrow$1 in the clumps by means of the detection equation (see \citealt{Hofner00} for more details). We use the \co\:and C$^{17}$O J$=$2$\rightarrow$1 APEX observations presented in \cite{Leurini11}, assuming the relative abundance, $\phi$, equal to 4 as found by \cite{Wouterloot08}\footnote{Assuming a galactocentric distance of the source equal to D$_{GL}=$7.4 kpc as in \cite{Leurini19}.}. 
Both transitions were observed in seven single-pointing observations, centered at the coordinates of the clumps. We we did not consider the data of clumps 9 and 10 (defined in \citealt{Leurini11}), because they are not part of the source, nor the data of the hot-core (i.e. clump-1), due to the saturation of the continuum observations (especially at 250 micron).\\
\indent The ratio between the peaks of the two CO isotopologues line intensities in each clump is then equal to: 

\begin{multline}\label{eq:ratio}
R_{{\rm C^{18}O},{\rm C^{17}O}}=\frac{T_{MB,{\rm C^{18}O}}}{T_{MB,{\rm C^{17}O}}}= \\ 
\frac{\eta_{\rm C^{18}O}[J(T_{{\rm ex},{\rm C^{18}O}},\nu_{\rm C^{18}O})-J(T_{bg}, \nu_{\rm C^{18}O})](1- e^{-\tau_{\rm C^{18}O}})}{\eta_{\rm C^{17}O}[J(T_{{\rm ex},{\rm C^{17}O}},\nu_{\rm C^{17}O})-J(T_{bg}, \nu_{\rm C^{17}O})](1- e^{-\tau_{\rm C^{17}O}})},
\end{multline}

\noindent
where the considered transition of \co\:and C$^{17}$O is the J$=$2$\rightarrow$1, at frequencies $\nu_{\rm C^{18}O}=219.5$ GHz and $\nu_{\rm C^{17}O}=224.7$ GHz, respectively; $\tau$ is the optical depth of the two considered CO isotopologues, with $\tau_{\rm C^{18}O}=\phi~\tau_{\rm C^{17}O}$.\\
\indent In eq.~(\ref{eq:ratio}) we assumed that the J$=$2$\rightarrow$1 transition of the two CO isotopologues correct under the same excitation conditions (i.e. $T_{{\rm ex},{\rm C^{17}O}} = T_{{\rm ex},{\rm C^{18}O}}$; e.g. \citealt{Martin&Barrett78,Ladd04}), are tracing the same volume of gas. This assumption has the consequence that the {\it filling factor}, $\eta$, is the same for both the transitions and therefore it can be eliminated from eq.~(\ref{eq:ratio}).\\
\indent We fitted a linear relation between $log_{10}$($\tau$) and $log_{10}$(N(\HH))\footnote{We used a $\tau$-N(\HH) relation and not a $\tau$-N(\co) one because N(\HH) is not affected by opacity (i.e. $\kappa_\nu$ correction already applied in Sect.~\ref{temperature_maps}) and the first relation has less scatter than the second.}, and then computed $C_\tau$ to correct N(\co) following the schematic diagram in Fig.~\ref{fig:scheme}. 
The best linear fit was computed with a least squares regression by extracting the C$^{17}$O and \co~peak fluxes 10$^4$ times (step 2 in Fig.~\ref{fig:scheme}). The basic assumption - see step 1a - was that the rms ($\varepsilon$) is normally distributed\footnote{This step was performed using the {\sf numpy.random.normal} function of NumPy (\citealt{Numpy06}) v1.14.}. For each clump, the $\tau$ value was calculated following eq.~(\ref{eq:ratio}) including the contribution of the rms and assuming a 10\% calibration uncertainty ($\sigma$), summed in quadrature to $\varepsilon$ (i.e. $T_{MB,{\rm C^{18}O}} = T^{\rm C^{18}O}_{MB} + \sqrt{\varepsilon^2 + \sigma^2}$ and the same for $T_{MB,{\rm C^{17}O}}$). We note that when applying this method to select different values of $T_{MB}$ in the \co\:and C$^{17}$O spectra, the errors do not change significantly if the distributions are built by 10$^{3}$ elements or more. Applying this procedure to each clump - steps 1(a-d) - we derived the linear fit $log_{10}(\tau)-log_{10}[{\rm N(H_2)}]$. We repeated this procedure $10^4$ times and we generated a cube of CO column density maps, one for each solution $log_{10}(\tau)-log_{10}[{\rm N(H_2)}]$ found. At the end of this procedure, a distribution of $10^{4}$ values of N(\co) has been associated to each pixel, used to compute the average value of N(\co) and its relative error bar in each pixel. We finally imposed an upper limit for the correction at $\tau=2$. This value would only be achieved in the saturated hot-core region and thus no pixel included in our analysis is affected by this condition. We then generated the opacity map from the N(\HH) map. The results of this procedure are shown in Fig.~\ref{fig:OPACITY} ($b$), together with the \co\:integrated intensity map - Fig.~\ref{fig:OPACITY} ($a$). We note that the estimated $\tau$ values range between $\sim 0.1$ and $\sim 1.9$ along the main body of G351.77-0.51.\\
\indent The final \co\:column density map was derived by including the opacity correction using the bestfit final $log_{10}(\tau)-log_{10}[{\rm N(H_2)}]$ relation - i.e. log$_{10}$($\tau_{\rm C^{18}O}$) $=$ 2.5log$_{10}$(N(\HH))-57.4. The opacity-corrected column density map of \co\:is shown in Fig.~\ref{fig:OPACITY} (c).
Our correction has increased the \co\:column densities by up to a factor of 2.3, and the remain almost constant in the branches. Over the whole structure, the column densities of \co\:range between 1 and 6$\times$10$^{16}$~cm$^{-2}$.\\

To summarize, the \co\:column density map was generated under the assumption of LTE following eq.~(\ref{eq:nc18o}). Possible saturation in the continuum data and opacity effects have been considered. The column density map was then used to evaluate the depletion map, as discussed in the next section.

\begin{figure*}
\centering
\includegraphics[width=1.8\columnwidth]{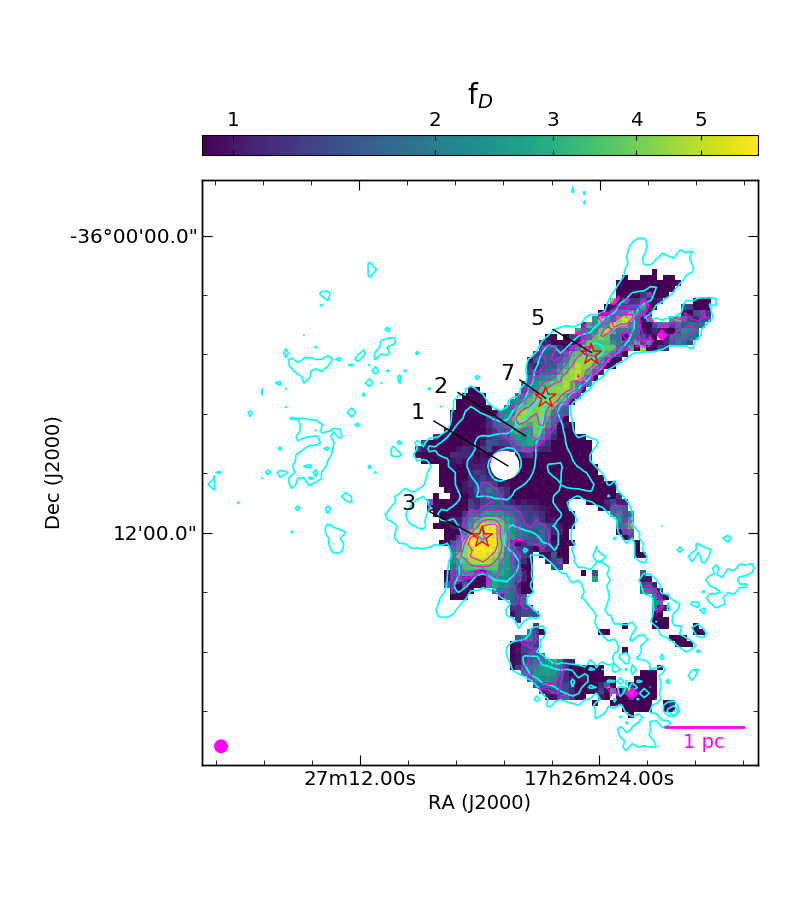}
\caption{Depletion factor ($f_D$) map obtained by taking the ratio between the expected and observed N(\co). We assumed a canonical abundance of 2.1$\times$10$^{-7}$ of N(\co) relative to N(\HH). The cyan contours are the same as in Fig.~\ref{fig:OPACITY}, while the ones in magenta are the defined at $f_D = $ 1.5, 2, 3, 4 and 5. Red stars in clump-5 and -7 indicate locations of the clumps reported in \citet{Leurini19}, while the one in clump-3 is set on the coordinates of the candidate HII region reported in \citet{Anderson14}.}
\label{fig:depletion_map}
\end{figure*}

\section{DISCUSSION}\label{discussion}
\subsection{The large-scale depletion map in G351.77-0.51}\label{observeddepletion}

\begin{figure}
\centering
\includegraphics[width=1.15\columnwidth]{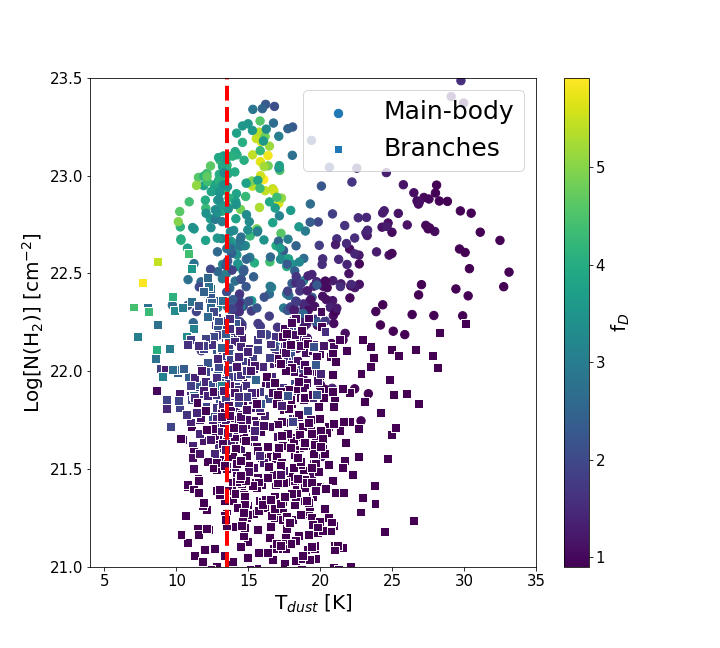}
\caption{Pixel-by-pixel scatter plot of the whole structure detected in Fig. \ref{fig:depletion_map}. Circles represent the pixels of the main body, while the squares indicate those of the branches. The red-dashed line represents the 10.8 K lower limit imposed on the $T_{\rm ex}^{\rm C^{18}O}$ (as discussed in Sect.~\ref{temperature_maps}) that corresponds to $T_{\rm dust}\sim 13.4$ K.}
\label{fig:scatter}
\end{figure}

The final CO-depletion factor map - Fig.~\ref{fig:depletion_map} - was generated by taking the ratio between the expected and the observed CO emission, using an abundance of \co\:with respect to \HH\:(i.e. $\chi^E_{\rm C^{18}O}$) equal to $2.1\times 10^{-7}$ (see \citealt{Giannetti17_oct}).\\
\indent As visible in Fig. \ref{fig:depletion_map}, in almost half of the map the depletion factor is $<$1.5, highlighting the absence of strong CO depletion around the core and along the branches in the southern directions with respect to the main ridge. This effect can have two different causes for the region of the central clump and for the branches, respectively: in the first case (i.e. the surroundings of clump-1 in Fig. \ref{fig:leurini}), the absence of depletion can be linked to the intense star formation activity, demonstrated in previous papers (e.g. \citealt{Leurini11}, \citealt{Konig17}, \citealt{Giannetti17_june} and \citealt{Leurini19}). The increase in temperature induced by the forming stars is able to completely desorb the ice mantles around dust grains in which the CO molecules are locked. This effect lowers the observed depletion factor, until it reaches unity. Following eq.~(2) in \cite{Schuller09}, we computed N(\HH) in the hot-core (saturated) region from the ATLASGAL peak flux at 870 $\mu$m, using $\kappa_{870 \mu{\rm m}} =$ 0.76 $\mathrm{cm^2\, g^{-1}}$ consistent with the dust opacity law discussed in Sect.~\ref{temperature_maps}. To obtain a depletion of 1 here, consistent with the surroundings, the dust temperature should be 80 K.\\
\indent To evaluate the effects of N(\HH) and T$_{dust}$ on $f_D$, Fig.~\ref{fig:scatter} was obtained by the pixel-by-pixel combination of Figures~\ref{fig:Tmaps} (b), \ref{fig:Tmaps} (a) and \ref{fig:depletion_map}. Each point represents a pixel shared between the three maps, where dots and squares are used to distinguish the pixels of the main body from those of the branches, respectively.\\
\indent Instead, in the branches (square markers in Fig. \ref{fig:scatter}) we notice that depletion reaches values $\sim$3.5. This result suggests that even in these structures the depletion process can start to occur. On the other hand, where $f_D$ is close to 1, the lower density disfavours a high degree of depletion (e.g. see \citealt{Caselli99}). 
Furthermore, we should consider that the observed depletion factor is averaged along the line-of-sight and in the beam.\\
\indent Along the main ridge and in the surroundings of clump-3 (Fig.~\ref{fig:leurini}) the depletion factor ranges between 1.5 and 6 and reaches its maximum in clump-3. Both regions appear in absorption at 8 $\mu$m in the \cite{Leurini11} maps, showing \HH~column densities of a few 10$^{22}$ cm$^{-2}$. In particular, it may appear counter-intuitive that we observe the highest depletion factor of the whole structure in clump-3, as an HII region has been identified in Wide-field Infrared Survey Explorer (WISE) at a distance of only $\approx 8''$ from the center of clump-3 (\citealt{Anderson14}). For these reasons, clump-3 should show similar depletion conditions to those in clump-1. However, such a high depletion factor suggests dense and cold gas close to the HII region contained in the clump. Within the cloud, the degree of depletion could be maintained if self-shielding would attenuate the effects of the radiation field of the HII region.
These ideas are supported by the analysis of the 8 and 24 $\mu$m maps shown in \cite{Leurini19}. At the location of the clump-3, a region slightly offset with respect to the bright spot associated with the HII region, is clearly visible in absorption at both wavelengths.\\
\indent In clumps 5 and 7 (i.e. regions C5 and C7 in Fig.~\ref{fig:leurini}), along the main ridge of G351.77-0.51, the average depletion factors are f$_{D, C7}=3.4^{+0.4}_{-0.5}$ (in clump-7; peaking at 4.5) and f$_{D, C5}=3.1^{+0.5}_{-0.6}$ (in clump-5; peaking at 4.3), respectively (see Sect.~\ref{Rfd_estime} for more details about error estimation). Compared to the average values of the samples mentioned in Sect.~\ref{sec_intro}, our values are slightly lower. %In this regard, 
However, we should consider that the observed depletion factors are affected by many factors such as the opacity correction applied and the \co/\HH~abundance ratio, which can vary up to a factor of 2.5 (e.g. \citealt{Ripple13}).\\
\indent Along the main ridge the depletion factors reach values of $\sim$ 6.
This is comparable to what is found in \cite{Hernandez11}, where the authors studied depletion factor variations along the filamentary structure of IRDC G035.30-00.33 with IRAM 30-m telescope observations. 
Along the filament, they estimate a depletion factor of up to 5. Of course, also in this case the considerations made earlier hold, but if we consider the different opacity corrections applied and \co/\HH~relative abundance assumed, we note the difference between our results and those of \cite{Hernandez11} are not larger than $\Delta f_D \sim$ 1.

To summarise, the final depletion map of G351.77-0.51 - Fig.~\ref{fig:depletion_map} - shows widespread CO-depletion in the main body, as well as at various locations in the branches. Comparing our results with those of \cite{Hernandez11} in G035.30-00.33, we note that in both cases the phenomenon of depletion affects not only the densest regions of the clumps, but also the filamentary structures that surround them. This result suggests that CO-depletion in high-mass star forming regions affects both small and large scales.

\subsection{Depletion modeling}\label{model}
\begin{figure}
\centering
\includegraphics[width=1\columnwidth]{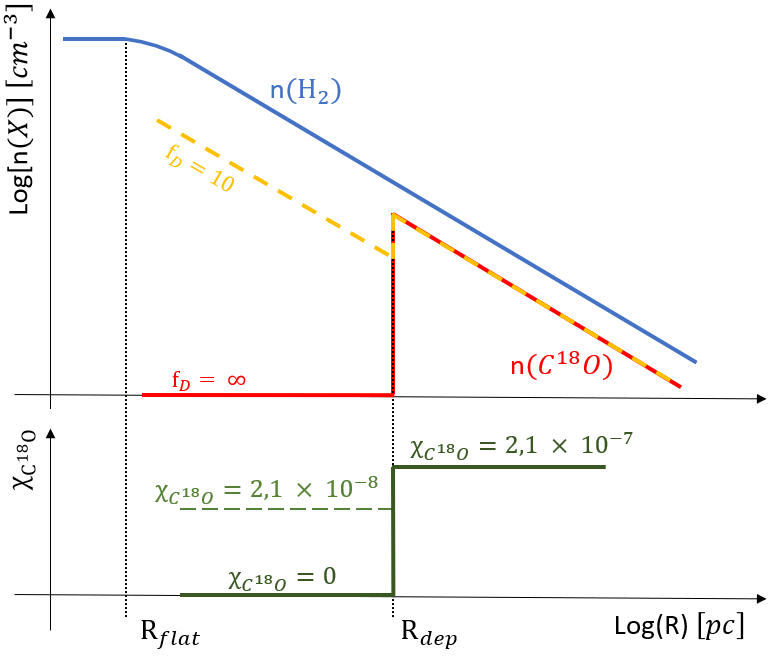}
\caption{Schematic representation of the model applied to simulate the depletion effect inside/outside the depletion radius. In blue we show the number density profile of \HH; in green the step-function reproducing the variation of the degree of depletion inside/outside $\Rd$. The C$^{18}$O profiles are obtained by the product of the other two.}
\label{fig:model}
\end{figure}

\begin{figure*}\label{profiles_clumps}
\centering
%\subfloat[][\emph{Region C5: $N$(\HH) profile}]
{\includegraphics[width=.322\textwidth]{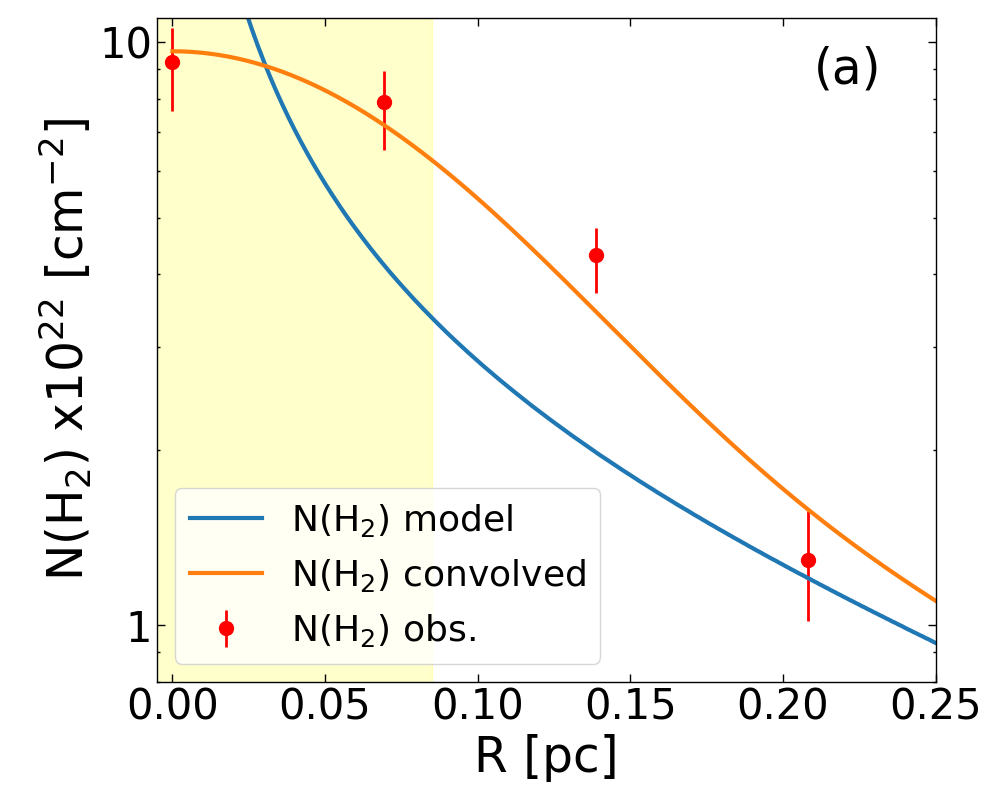}} 
%\subfloat[][\emph{Region C5: $N$(\co) profile}]
{\includegraphics[width=.322\textwidth]{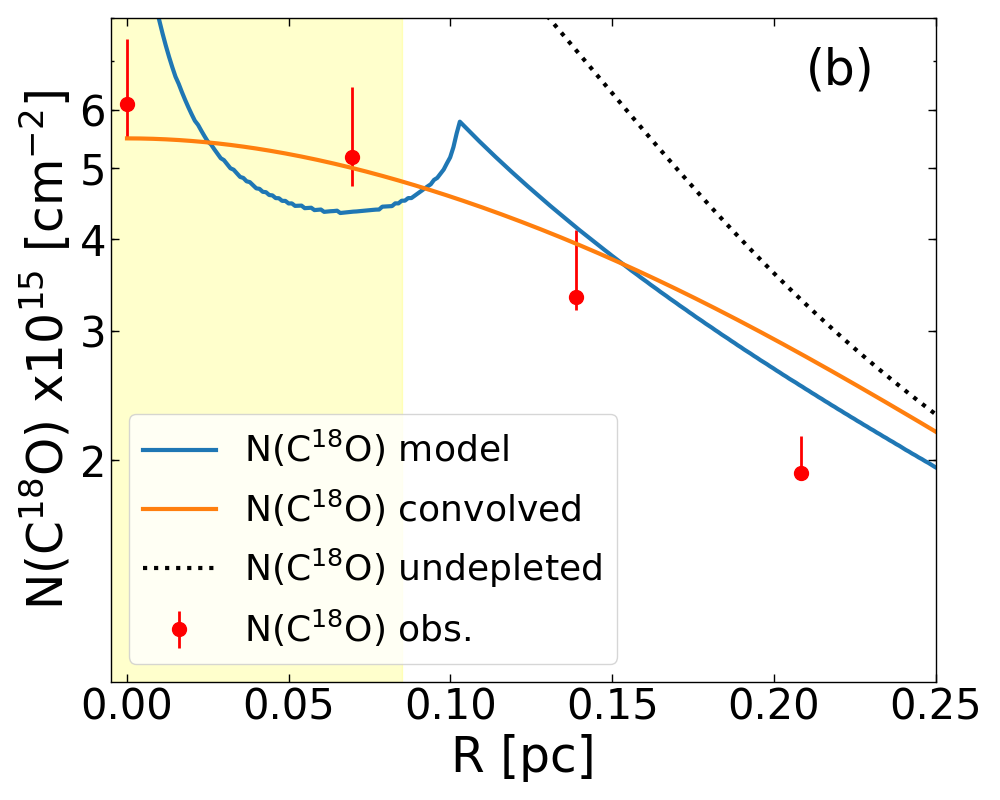}}
%\subfloat[][\emph{Region C5: $f_D$ profile}]
{\includegraphics[width=.322\textwidth]{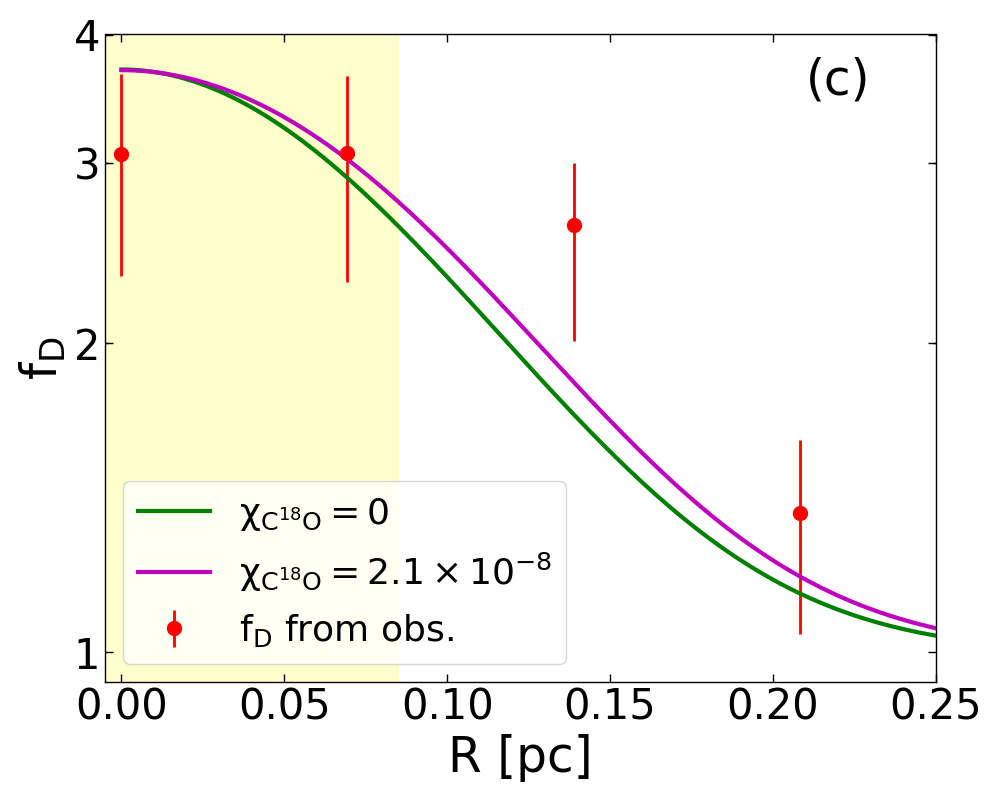}}\\
%\subfloat[][\emph{Region C7: $N$(\HH) profile}]
{\includegraphics[width=.322\textwidth]{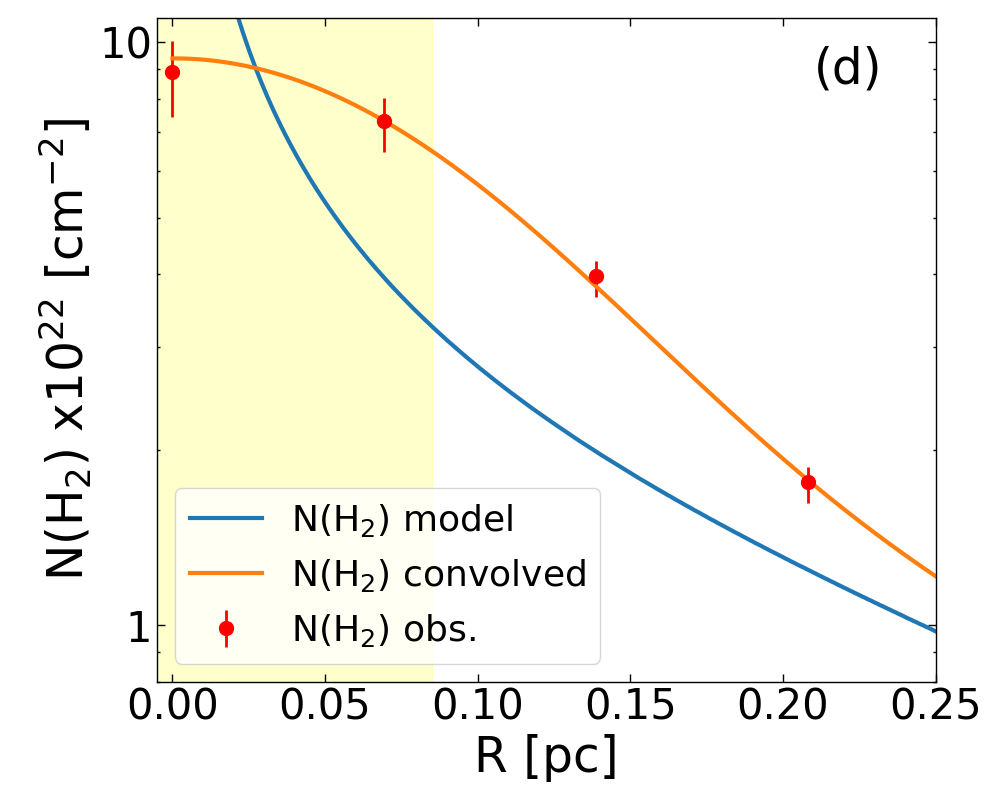}} 
%\subfloat[][\emph{Region C7: $N$(\co) profile}]
{\includegraphics[width=.322\textwidth]{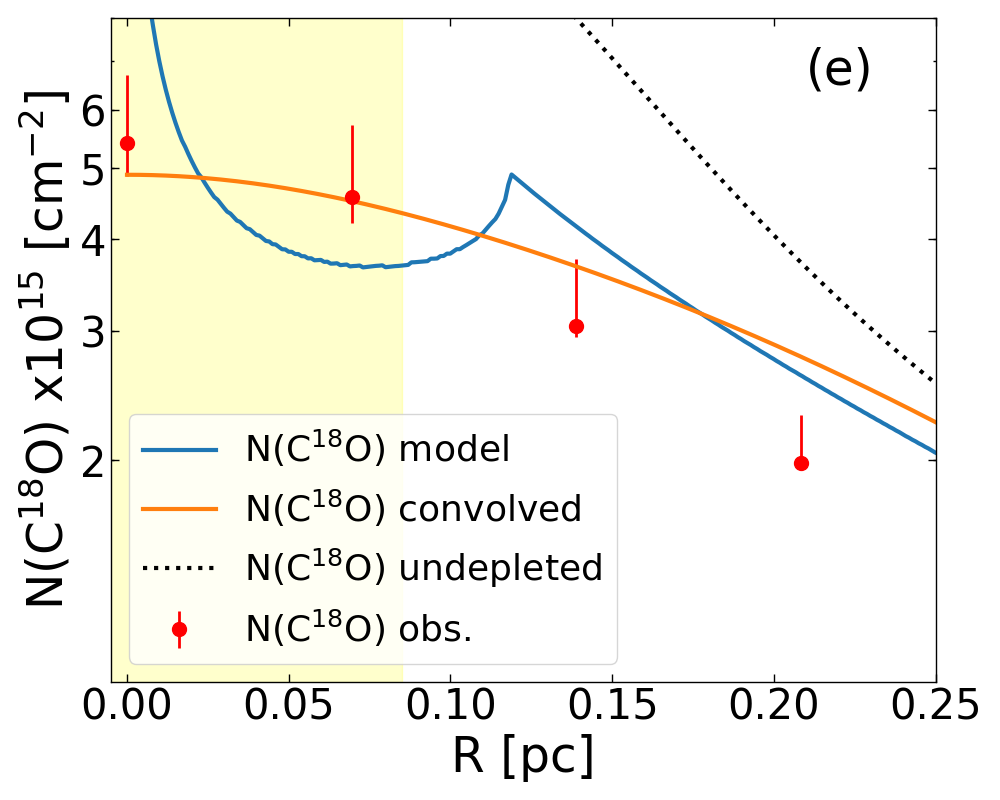}}
%\subfloat[][\emph{Region C7: $f_D$ profile}]
{\includegraphics[width=.322\textwidth]{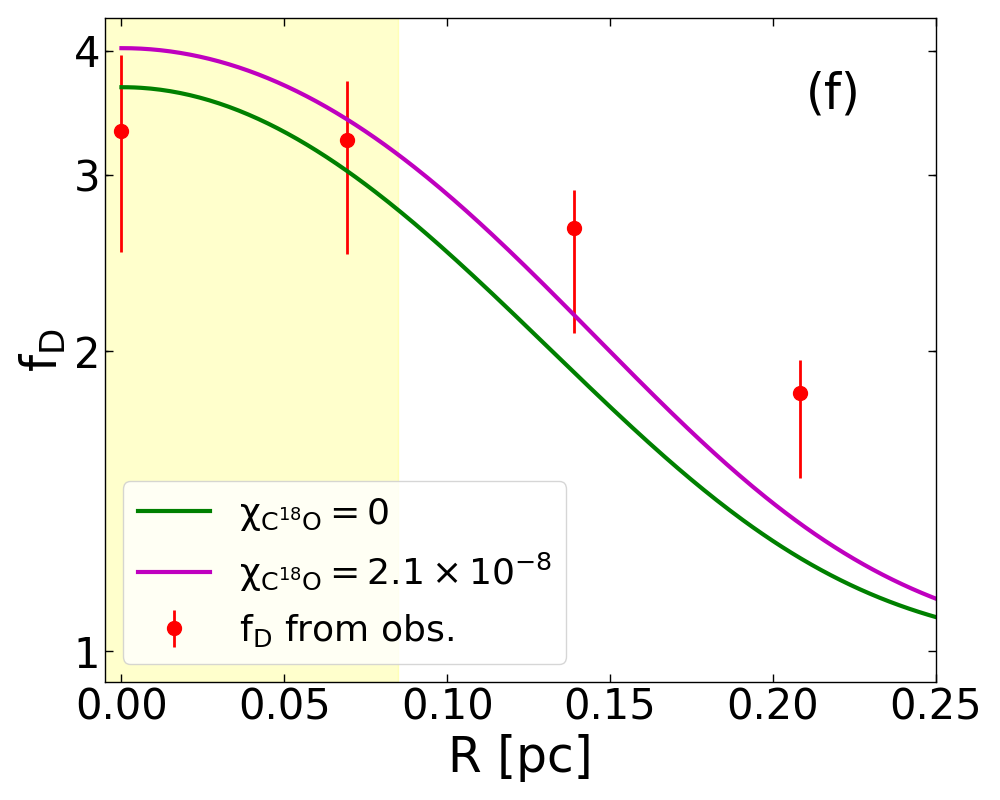}}\\
%\subfloat[][\emph{Region F1: $N$(\HH) profile}]
{\includegraphics[width=.322\textwidth]{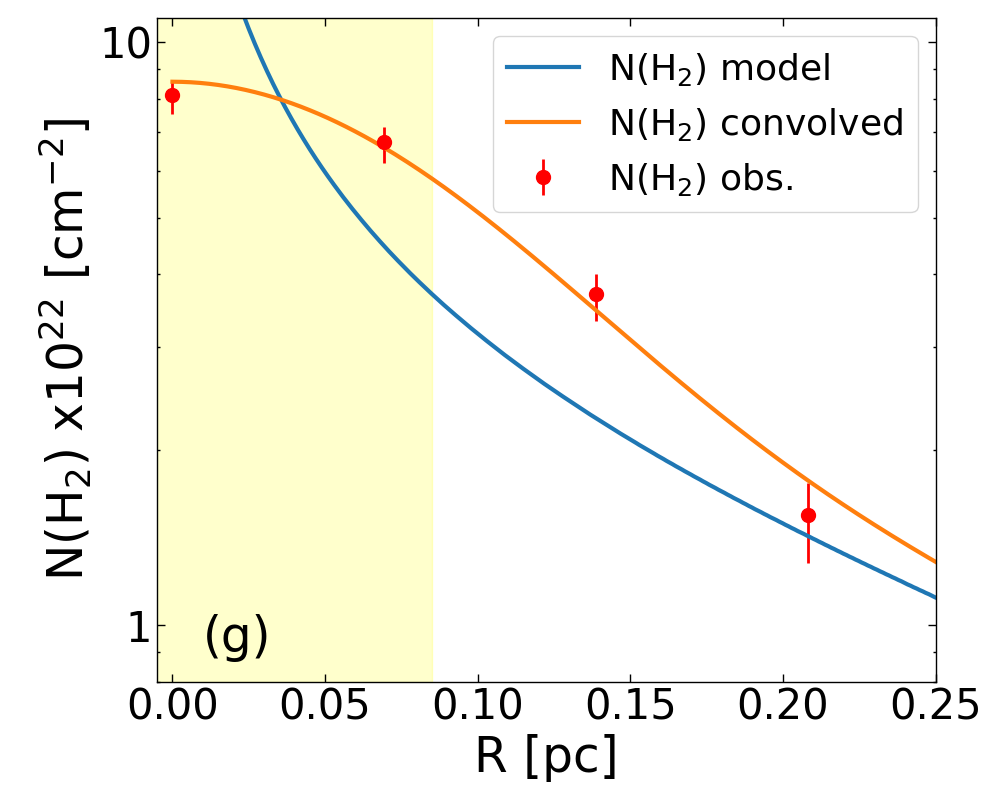}} 
%\subfloat[][\emph{Region F1: $N$(\co) profile}]
{\includegraphics[width=.322\textwidth]{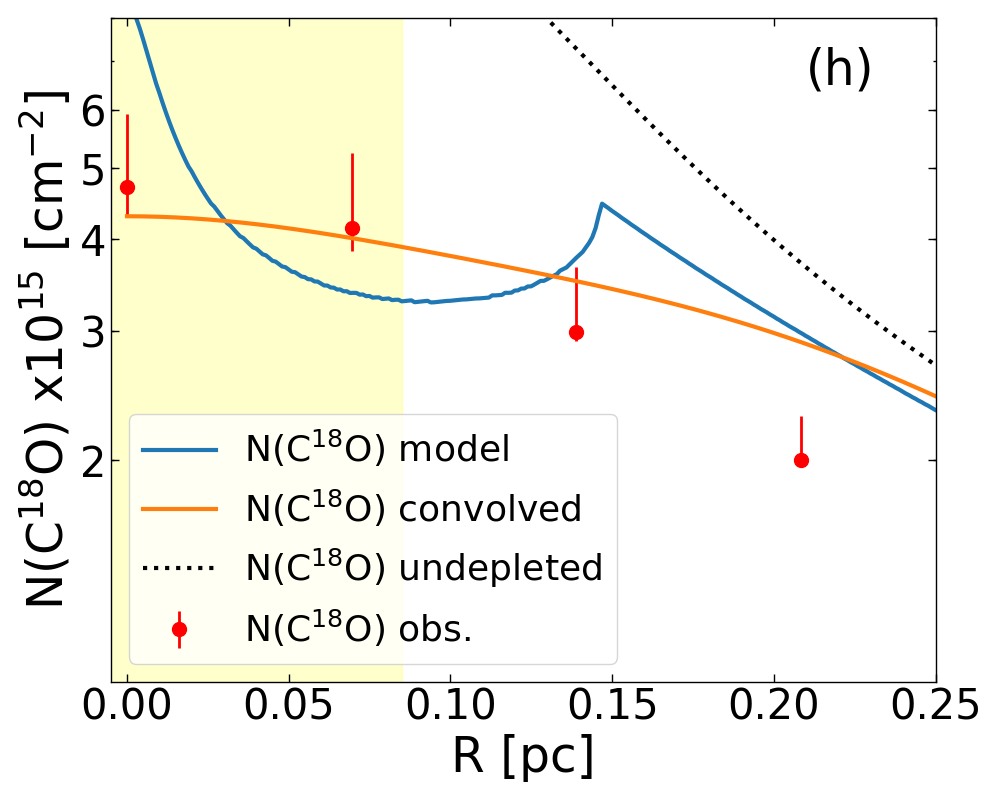}}
%\subfloat[][\emph{Region F1: $f_D$ profile}]
{\includegraphics[width=.322\textwidth]{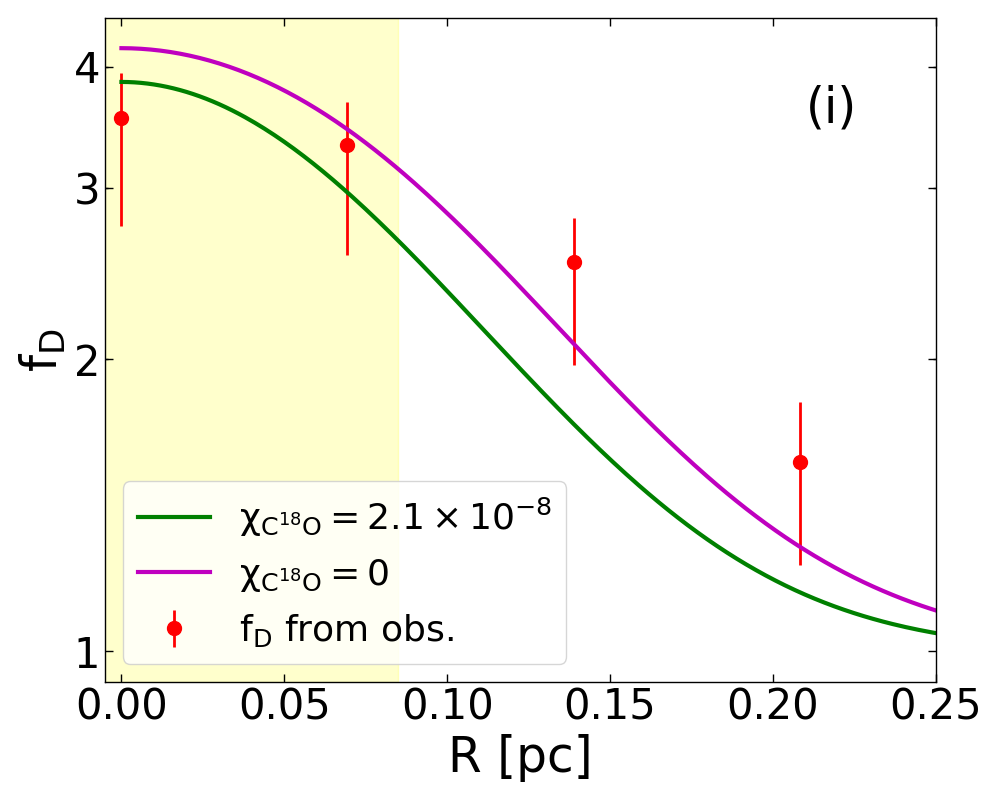}}

\caption{From top: Column density profiles of \HH\:and \co\:in panel (a) and (b), respectively, while panel (c) shows the los- and beam-averaged depletion factor profile from region C5 in Fig. \ref{fig:leurini}. Panels (d), (e) and (f) are the same as (a), (b) and (c), respectively, for region C7, while panels (g), (h) and (i) refer to the region F1. Blue profiles in panels (a), (d) and (g) are the synthetic column densities profiles obtained by the integration of the number densities distribution; solid-blue lines in panels (b), (e) and (h) are the same for \co\:assuming $\chi_{\rm C^{18}O} = 2.1 \times 10^{-8}$ for R$<\Rd$, and the estimated position of $\Rd$ corresponds to the cusp. Orange profiles in panel (a), (b), (d), (e), (g) and (h) are the convolution results of the blue profiles and the {\it Herschel} beam at 500$\mu$m, while black-dotted line in panel (b), (e) and (h) indicates the convolved undepleted-profile of N(\co) assuming a constant $\chi_{\rm C^{18}O} = 2.1 \times 10^{-7}$. Green and magenta profiles in panel (c), (f) and (i) show the depletion factor profiles obtained by assuming different abundances for R$<\Rd$ (see eq.~\ref{eq:stepfunction}). Red points are the observed values for each quantity plotted. The yellow- shaded area shows the dimensions of the {\it Herschel} HWHM at $\lambda= 500~\mu$m (i.e. FWHM$= 36\arcsec$; 0.17 pc at the source distance). All profiles are plotted as a function of the radial distance from the spine.}
\label{fig:clumpsProfiles}
\end{figure*}

High densities and low gas temperatures favour CO-depletion. Given this, it is reasonable to think that \co/\HH~is not constant within a cloud. This quantity varies as a function of location, following the volume density and gas temperature distributions/profiles.\\
\indent The regions in which the depletion degree is higher are those where the dust surface chemistry becomes more efficient, due to the high concentration of frozen chemical species on the dust surface. To understand how the efficiency of the various types of chemical reactions change, we need to understand how the depletion degree varies within the dark molecular clouds.\\
\indent In order to reproduce the average depletion factor observed in G351.77-0.51 we built a simple 1D model describing the distributions of \co\:and \HH.
We focus our attention on the main ridge, identifying three distinct regions: clumps 5 and 7, and the filamentary region between them (i.e. regions C5, C7 and F1 in Fig.~\ref{fig:leurini}, respectively).\\
\indent The model assumes that both profiles have radial symmetry with respect to the center of the ridge, i.e. the ``spine''. $R_{flat}$ is the distance relative to the spine within which the density profiles remain roughly flat (i.e. $n({\rm H}_2, R_{flat})=0.5n( {\rm H}_{2,spine})$ if $p=2$) . We normalize the volume density profiles of the model with respect to R$_{flat}$, while $n({\rm H}_{2,spine})$ is the central volume density of \HH.\\
\indent Our \HH~volume density profile is described by:

\begin{equation}\label{eq:nH2model}
n({\rm H}_2) = n( {\rm H}_{2,spine}) \left[1+\left(\frac{R}{R_{flat}} \right)^{\alpha} \right]^{-p/2}, 
\end{equation}

\noindent 
up to a maximum distance R$_{max}\sim 0.2$ pc, two times larger than the filament width estimated in \cite{Leurini19}, that encompasses the entire filamentary structure.\\
\indent For the clumps (i.e. regions C5 and C7), we assume $p=2$ and eq.~(\ref{eq:nH2model}) takes the functional form described by \cite{Tafalla02}. In this case, the free parameters of the model are $\alpha$, $n({\rm H}_{2,spine})$ and R$_{flat}$. 
Starting from R$_{flat}$, both the distributions of \co~and \HH~scale as a power-law with the same index: $\alpha_{C5}=1.9$ and $\alpha_{C7}=1.8$, for the clump-5 and clump-7, respectively. The solutions for $\alpha$ was found by exploring the parameter space defined by the results of \cite{Beuther02}, i.e. 1.1$<\alpha<$2.1. The best-fitting models returns the values of $n({\rm H}_{2,spine})$ and R$_{flat}$ equal to 8$\times 10^6$ cm$^{-3}$ and 5.5$\times$10$^{-3}$ pc in the case of clump-5 and 5.8$\times 10^7$ cm$^{-3}$ and 2$\times$10$^{-3}$ pc for the clump-7.\\
\indent In the case of the main body (i.e. region F1), $\alpha = 2$ the volume density profile of \HH\:is described by a Plummer-like profile (see \citealt{Plummer11} for a detailed discussion). For this model, the free parameters are $p$ and $n({\rm H}_{2,spine})$, while R$_{flat}$ corresponds to the thermal Jeans length, $\lambda_{J}$, for an isothermal filament in hydrostatic equilibrium, calculated at R$=0$:

\begin{equation}\label{eq:lambdaJ}
\lambda_{J}=\frac{c^2_s}{G~\mu~m_H~{\rm N}({\rm H}_2)_{R=0}},
\end{equation}

\noindent
where $c^2_s$ is the isothermal sound speed for the mean temperature in the region F1 (equal to 13.8 K), $G$ is the gravitational constant in units of [cm$^3$ g$^{-1}$ s$^{-2}$], $\mu$ is the mean molecular weight of interstellar gas ($\mu=2.3$), $m_H$ is the hydrogen mass in grams and N$({\rm H}_2)_{R=0}$ is the averaged \HH~column density calculated at R$=0$. Applying eq.~(\ref{eq:lambdaJ}) to region F1, we obtained R$_{flat} \sim 0.008$ pc, while the value of $p$ index was found equal to $1.9$ after exploring the range of values reported in \cite{Arzoumanian11}. We note that an R$_{flat}$ variation of $~30\%$ does not significantly change the N(H$_2$) profile.\\
\indent To simulate depletion effects, we introduced the depletion radius, $\Rd$, that is the distance from the spine where the degree of depletion drastically changes, following equations:

\begin{equation}\label{eq:stepfunction}
 \left. 
 \begin{aligned}
 & R\leqslant\Rd \qquad f_D = (10, \infty); \qquad  \chi_{\rm C^{18}O} = \frac{\chi^E_{\rm C^{18}O}}{f_{D}}\\
 & R>\Rd \qquad f_D = 1; \qquad  \chi_{\rm C^{18}O} = \chi^E_{\rm C^{18}O} 
    \end{aligned}
 \right\},
 \end{equation}

\noindent
where the canonical abundance for \co\:($\chi^E_{\rm C^{18}O}= 2.1 \times 10^{-7} $) was set by using the findings of \cite{Giannetti17_oct} and references therein. The lower limit of $f_D=10$ for$R<\Rd$, is motivated by the observational constraints for high-mass clumps, which in most cases show depletion values between 1 and 10 (e.g. \citealt{Thomas08, Fontani12, Giannetti14}, as discussed in Sect.~\ref{sec_intro}). Instead, the choice of extending the model until reaching the theoretical condition of full depletion (i.e., $f_D=\infty$) is done to study the effect that such a drastic variation of $f_D$ has on the size of $\Rd$.\\
\indent In Fig.~\ref{fig:model} we present a sketch of the model: the profile for $n$(\HH) is indicated in blue, that for $\chi_{\rm C^{18}O}$ is shown in green (bottom panel), while profiles for $n$(\co) appear in orange and red (two values of $f_D$ are considered). Radial profiles are then convolved with the {\it Herschel} beam at $\lambda= 500~\mu$m (i.e. FWHM$= 36\arcsec$). For each profile we calculate the set of values that best represent the observed data in each considered region (i.e. C5, C7 and F1 of Fig.~\ref{fig:leurini}): we generate a set of profiles of $n$(\HH) by exploring the parameter space defined by the free parameters of each modeled region, and we select the curve that maximizes the total probability of the fit (i.e. P$_{\rm TOT}={\prod_{i} p(r_i)}$, where $p(r_i)$ are the probability density values corresponding to the error bars in Fig.~\ref{fig:clumpsProfiles}, and calculated at the various distances - $r_i$ - from the spine). The same was done for the profile of $n$(\co), selecting the best profile from a set of synthetic profiles generated by assuming different sizes of $\Rd$.\\

In the next section, we present the results of the model applied to the clumps and the filament region, providing an estimate for both $\Rd$ and $n$(\HH) calculated at $\Rd$.

\begin{figure}
\centering
\includegraphics[width=1\columnwidth]{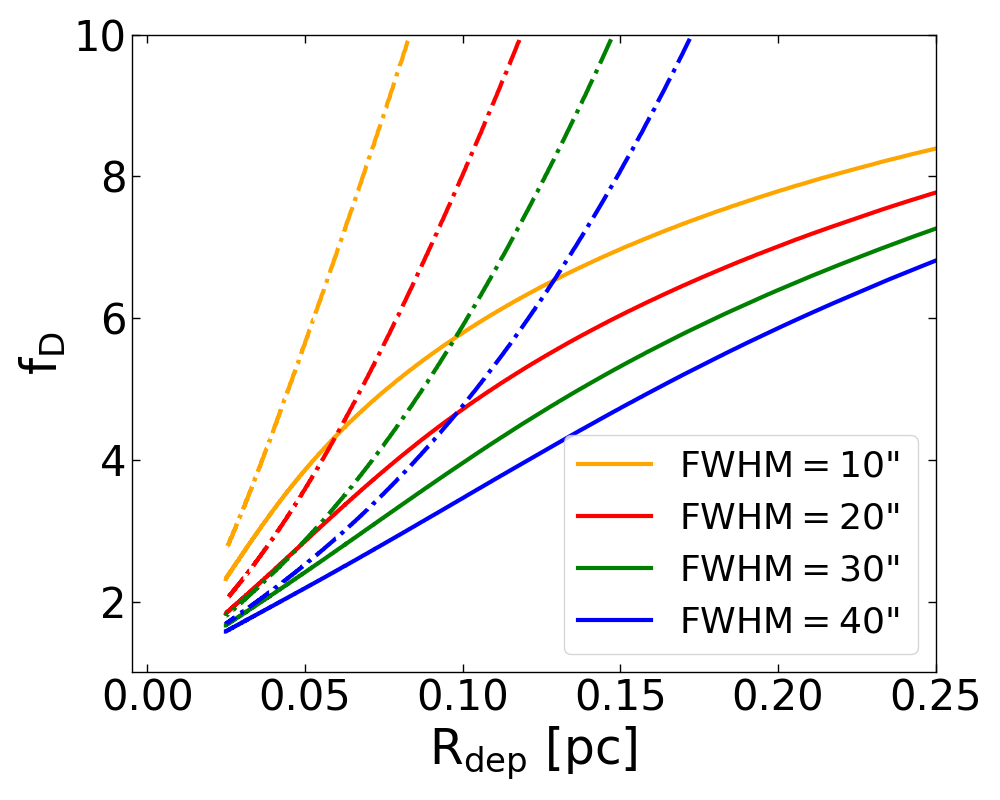}
\caption{$f_D$ variation as a function of $\Rd$ for the best-fitting model applied to clump-5. Full lines represent the $f_D$-$\Rd$ relation assuming $f_D=$ 10 for R$<\Rd$, while the dash-dotted line are the ones for $f_D=\infty$. The blue profile in Fig.~\ref{fig:clumpsProfiles} (a) was convolved at different angular resolutions in order to predict how the same relation changes with different observational conditions. Colours are linked to the different FWHMs used for the convolution.}
\label{fig:fDRdep_relation}
\end{figure}

\subsubsection{Estimate of $\Rd$}\label{Rfd_estime}
\noindent
In Fig.~\ref{fig:clumpsProfiles} ($a$) and ($b$) we show the best-fitting mean-radial column density profiles of N(\HH) and N(\co) of clump-5, respectively. In blue, we show the column density profiles before the convolution while in orange the convolved profiles. Black-dashed line in the central panels are the convolved undepleted-profile of N(\co), assuming a constant $\chi_{\rm C^{18}O} = 2.1 \times 10^{-7}$ for all R. The observed data are plotted as red points. In Fig.~\ref{fig:clumpsProfiles} ($c$) we report (in green) the depletion factor as a function of the distance from the projected centre in clump-5 assuming an abundance of \co\:for R$<\Rd$ equal to 2.1$\times 10^{-8}$, while in magenta the same profile for $\chi_{\rm C^{18}O} = 0$. Figures~\ref{fig:clumpsProfiles} ($d, e, f$) are the same for the clump-7, respectively. In both cases, we estimate $\Rd$ by varying the abundance $\chi_{\rm C^{18}O}$ for $R<\Rd$ considering the two limiting cases, i.e. with $f_D =$ 10 and $f_D = \infty$.\\
\indent All error bars have been defined taking the range between the 5$th$ and 95$th$ percentiles. In the case of N(\HH), the uncertainties are generated by computing the pixel-to-pixel standard deviation of the \HH~column density along the direction of the spine, and assuming a Gaussian distribution of this quantity around the mean value of N(\HH). The errors associated with N(\co) are computed propagating the uncertainty on the line integrated intensity and on the optical depth correction (as discussed in Sect.~\ref{opacity}), using a Monte Carlo approach. Finally, the uncertainties on $f_D$ are estimated computing the abundance, randomly sampling the probability distributions of N(\HH) and N(\co).\\
\indent In clump-5, the estimated $R_{dep,i}$, ranges between a minimum of $\sim$ 0.07 pc and a maximum of $\sim$ 0.10 pc. In clump-7 instead, which shows a larger average depletion factor compared to clump-5, $\Rd$ ranges between $\sim$ 0.07 and 0.12 pc (for $f_D=\infty$ and $f_D=$10, respectively). Comparing Fig.~\ref{fig:clumpsProfiles} (c) and (f), both synthetic profiles are able to well reproduce the observed ones up to a distance of $\sim$0.15 pc.
In the case of total depletion regime (i.e. $f_D=\infty$), at $\Rd$ we estimate a volume density $n$(\HH)$ = 5.7 \times 10^4$ cm$^{-3}$ and $5.3 \times 10^4$ cm$^{-3}$, for clump-5 and 7, respectively. Our results are comparable with those found in low-mass prestellar cores (i.e. \citealt{Caselli02, Ford-Shirley11}) and in high-mass clumps in early stages of evolution (i.e. \citealt{Giannetti14}). Furthermore, it is important to note that both values should be considered as upper limits due to the approximation in eq.~(\ref{eq:stepfunction}) for all the models discussed. For the case in which $f_D = 10$, in clump-5 at a distance equal to $\Rd = 0.10$ pc, the model provides a volume density of $n$(\HH)$ = 3.1 \times 10^4$ cm$^{-3}$. Similarly, in the case of clump-7, $\Rd = 0.12$ pc is reached at a density of $n$(\HH)$ = 2.2 \times 10^4$ cm$^{-3}$.\\
\indent In Fig.~\ref{fig:clumpsProfiles} ($g$) and ($h$) we show the mean-radial column density profiles of N(\HH) and N(\co), respectively, in the region between the two clumps (i.e. F1 region in Fig.~\ref{fig:leurini}), while the depletion profile is shown in the panel ($i$). The results are in agreement with those estimated for clumps 5 and 7, showing a 0.08 pc $<\Rd<$ 0.15 pc (comparable with the typical filament width $\sim$ 0.1 pc; e.g. \citealt{Arzoumanian11}), which corresponds to densities of $4.2 \times 10^4$ cm$^{-3}$ $>$ $n$(\HH) $>$ 2.3 $\times 10^4$ cm$^{-3}$ for $f_D=\infty$ and 10 respectively. The summary of the $\Rd$ estimates is shown in Tab.~\ref{tab:Rdep} (Sect.\ref{caveats}).\\
\indent Once the model presented in Sect.~\ref{model} has been calibrated with the data, it allows one to estimate how the size of $\Rd$ can vary according to the obtained $f_D$. An example of these predictions is shown in Fig.~\ref{fig:fDRdep_relation}. Starting from the best fitting synthetic profile obtained for clump-5 (i.e. the blue profile in Fig.~\ref{fig:clumpsProfiles}a), we generated a family of N(\co) profiles, convolved with different beams (i.e. FWHM between 10 and $40''$). The value of $f_D$ reported in Fig.~\ref{fig:fDRdep_relation} is calculated at $R=$ 0 according to eq.~(\ref{eq:depletionfactor}), by varying the dimensions of $\Rd$ from 0 up to the maximum size of the model radius for both $f_D$ = (10, $\infty$) at$R<\Rd$ (i.e. full and dash-dotted lines, respectively).￼ We stress that this type of prediction is strongly dependent on the model assumptions (i.e. the assumed \co/\HH~abundance) and valid only for sources for which clump-5 is representative.\\
Nevertheless, a variation in $\Rd$ is notable and it changes by varying the value of $f_D$ for $R<\Rd$: at FWHM $=$ 10'' for example, a factor of 1.5 in $f_D$ (from 4 to 6) corresponds to a factor of 1.6 in the $\Rd$ size if we assume $f_D=\infty$ for $R<\Rd$, while for $f_D=$ 10 it corresponds to a factor of 2.2. These variations are slightly lowered if FWHM increases: a factor of 1.4 and 1.6 for $f_D$ = ($\infty$, 10) at $R<\Rd$, respectively for FWHM $=$ 40''.

\subsubsection{Comparison with 3D models}\label{Rfd_in_C1C2}
As a first step, we compared the estimated values of $\Rd$ with the results obtained by \cite{Koertgen17}. These authors imposed a condition of total depletion on the whole collapsing region of their 3D MHD simulations, to model the deuteration process in prestellar cores. The initial core radius range is 0.08 $<R_c<$ 0.17 pc, depending on the simulation setup. This assumption is necessary to reduce the computational costs of the 3D-simulations but, as discussed in Sect.~\ref{sec_intro}, the outcome can strongly deviate from reality when we move to larger scales.\\
\indent For G351.77-0.51, the largest estimated depletion radius is $\lesssim 0.15$ pc, comparable with the initial core size assumed by \cite{Koertgen17}.\\ 
\indent We note also that, imposing the condition of total depletion on a scale of 0.17 pc, \cite{Koertgen17} find a deuterium fraction of \HH D$^+$ reaching value of 10$^{-3}$ (100 times the canonical values reported in \citealt{Oliveira03}) on scales comparable with our size of $\Rd$, after $\sim$ 4$\times$10$^4$ yr (see their Fig. 3). This result is in agreement with the ages suggested by the depletion timescale estimated in our models following \cite{Caselli99}:

\begin{equation}\label{depletion_time}
  \tau_{dep} \sim \frac{10^9}{S~n(H_2)}\,\,\, \mathrm{yr},
\end{equation}

\noindent
assuming the estimated volume densities at $\Rd$ and a sticking coefficient S$=$1.\\
\indent A similar picture is reported by \cite{Koertgen18}, who show that the observed high deuterium fraction in prestellar cores can be readily reproduced in simulations of turbulent magnetized filaments. After $\sim$10$^4$ yr, the central region of the filament shows a deuterium fraction 100 times higher than the canonical value, and has a radius of $\sim$0.1~pc (see their Fig.~1). The dimensions of the regions showing a high deuterium fraction can be compared with our estimates of $\Rd$ since the two processes are connected, as shown in Sect.~\ref{sec_intro}.

The comparison of our findings with both small- and large-scale simulations, suggests that the total-depletion assumption could provide reliable results within reasonable computational times.

\section{The influence of a different abundance profile on $\Rd$}\label{caveats}

\begin{figure}
\centering
\includegraphics[width=1\columnwidth]{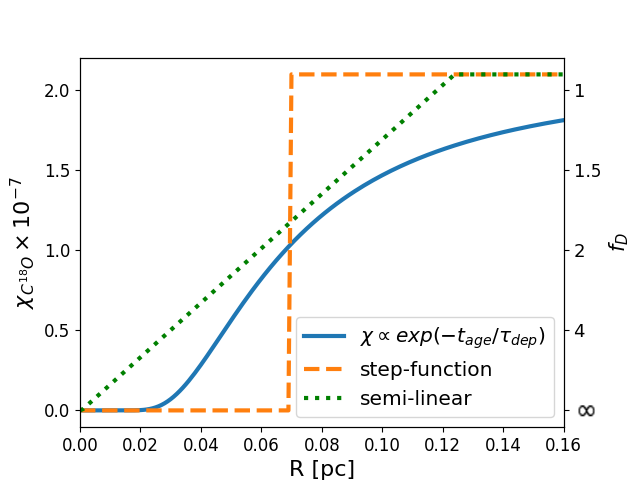}
\caption{Sketch of the abundance profiles assumed to describe the $\chi_{\rm C^{18}O}$ variation in our tests. The orange-dashed profile shows the step function described in the Sect.~\ref{model}. In this plot, only the position of the step-function discontinuity has a physical meaning, equal to the estimated size of $\Rd$ in the C5 model with $\chi_{\rm C^{18}O}=0$ for R$<\Rd$; the green-dotted profile describes a more gradual variation of $\chi_{\rm C^{18}O}$ for R$<R_{drop}$ as a linear profile between canonical \co/\HH~abundance and the value of $\chi_{\rm C^{18}O}$ at the center of the model; blue profile reproduces the trend $\chi_{\rm C^{18}O} \propto exp(-t_{age}/\tau_{dep})$, assuming only the CO absorption as dominant process (see Sect.\ref{caveats}).}
\label{fig:XIprof}
\end{figure}

It is important to stress that the model discussed in this paper presents a number of limitations:
\begin{itemize}
\item A first caveat is represented by how we obtain the final values from observations (i.e. red dots in Fig.~\ref{fig:clumpsProfiles}), with which our models were then optimized. The main assumption we made was the canonical \co/\HH\:abundance, assumed equal to $2.1\times 10^{-7}$, which actually can vary by up to a factor of 2.5 (Sect.~\ref{observeddepletion}). While knowing these limits, we note that the results shown in the final $f_D$ map of G351.77-0.51 (i.e. Fig.~\ref{fig:depletion_map}) are in good agreement with what was found by \cite{Hernandez11, Hernandez12} in the IRDC G035.30-00.33, the only other case in which a filamentary-scale depletion factor map has been shown to date;
\item A second limitation comes from the \HH~volume density profile we assumed in Sect.~\ref{model} to reproduce the data, which ignores all the sub-structure of clumps. Although this assumption is strong, it seems reasonable because the best fitted N(\HH) profiles shown in Fig.~\ref{fig:clumpsProfiles} - panels (a), (d) and (g) - are in agreement with the data points;
\item The third and strongest assumption we made is the shape of the $\chi_{\rm C^{18}O}$ profile to describe the \co~variation with respect to \HH~within the main ridge of G351.77-0.51. In Fig.~\ref{fig:XIprof} the step-function profile assumed in Sect.~\ref{model} is shown as an orange-dashed line. All the results of Figures~\ref{fig:clumpsProfiles} and \ref{fig:fDRdep_relation}, and discussed in Sections~\ref{Rfd_estime} and \ref{Rfd_in_C1C2}, assume this abundance profile.\\
\indent We therefore examined how much the estimates of the $\Rd$ size, reported in the previous paragraphs, depend on the abundance profile assumed. For these tests, we studied the three $\chi_{\rm C^{18}O}$ profiles shown in Fig.~\ref{fig:XIprof}. 
In addition to the already discussed step function, we tested two profiles describing a smoother variation of $\chi_{\rm C^{18}O}$ as a function of R. In the first test, the \co/\HH~abundance was assumed to be constant up to a certain distance, called $R_{drop}$, at which the abundance starts to linearly decrease (the green-dotted profile in Fig.~\ref{fig:XIprof}). In the second test, we considered the physical case in which $\chi_{\rm C^{18}O} \propto exp(-t_{age}/\tau_{dep})$, where $t_{age}$ is the age of the source, as expected integrating the evolution of $n($\co$)$ over time, $dn$(\co)$/dt$, only considering depletion (i.e. $dn$(\co)$/dt = -\kappa_{ads} n$(\co), and ignoring desorption). The latter case is shown as a blue profile in Fig.~\ref{fig:XIprof}, which reaches zero in the innermost regions of the model (i.e., in this example, for R $\lesssim 0.02$ pc), due to the high volume densities of \HH~(see eq.~\ref{depletion_time}).\\
\indent Since we no longer have a strong discontinuity in the new abundance profiles, we have to redefine the concept of $\Rd$. In the new tests, $\Rd$ will correspond to the distance at which $\chi_{\rm C^{18}O} = 2.1 \times 10^{-8}$ (i.e. the distance at which the 90\% of the CO molecules are depleted) and its new estimates are shown in Tab.~\ref{tab:Rdep}.\\
\indent We note that the exponential profile describes the physically more realistic case, whereas the other two are simpler approximations, easier to model. Therefore, we compare the results of the exponential case with those described by the other two.\\
\indent As visible in Tab.~\ref{tab:Rdep}, the new $\Rd$ estimated from the model with the exponential profile are within a factor of about 2-3 of those found from the semi-linear and the step-function models, respectively. These results suggest that, although the assumption of a step-function to describe the $\chi_{\rm C^{18}O}$ profile may seem too simplistic, it actually succeeds reasonably well in reproducing the results predicted by a physically more realistic profile. Furthermore, the step-function and semi-linear profiles represent two extreme (and opposites) cases of a drastic and a constant change in the $\chi_{\rm C^{18}O}$ profile, respectively, while the exponential profile can be interpreted as a physically more realistic situation. Because of this, even if the relative 3$\sigma$ level errors on $\Rd$, calculated by probability distributions, are always less than 15\% of the values reported in Tab.~\ref{tab:Rdep}, the uncertainty of a factor of 2-3 on the estimates of $\Rd$ from the exponential profile seems more realistic.

\begin{table}
 \centering
 \renewcommand\arraystretch{1.2}
 \caption{Summary of the estimated $\Rd$ assuming the profiles in Fig.~\ref{fig:XIprof} and discussed in Sect.~\ref{caveats}. For all the estimates of $\Rd$ shown in the table, we estimate the relative 3$\sigma$ level error to be less than 15\%.}\label{tab:Rdep}
 \begin{tabular}{c|p{1cm}|p{1cm}|p{1cm}}
 \hline
 \hline
	{\bf Regions}				& {\bf C5} 	& {\bf C7}	& {\bf F1} 	\\
	{\it Profiles}				&{\bf [pc]}	&{\bf [pc]}	&{\bf [pc]}	\\
 \hline
 {\it Step-function}			&	$0.10$	&	$0.12$	&	$0.15$	\\
  ($f_D = 10;$ R$<\Rd$)		&			&			&			\\ 
 {\it Step-function}			&	$0.07$	&	$0.07$	&	$0.08$	\\ 
  ($f_D = \infty;$ R$<\Rd$)	&			&			&			\\ 
 {\it Exponential}				&	$0.04$	&	$0.04$	&	$0.05$	\\
 {\it Semi-linear}				&	$0.02$	&	$0.02$	& 	$0.03$	\\ 
 \hline	
 \hline
\end{tabular}

\end{table}

\indent Finally, it is worth discussing the predictions of the depletion timescales in the three models. In clump-5, for example, the $\Rd$ of the model with the semi-linear profile, provides $n($\HH$) = 5.5 \times 10^{5}$ cm$^{-3}$. With this result, we estimated $\tau_{dep} = 1.8 \times 10^{3}$ yr, smaller by a factor of about 10-20 than those obtained using the model with the step-function profile, for which $\tau_{dep} = 1.8 \times 10^{4}$ yr ($f_D=\infty$ for R$<\Rd$) or $3.2 \times 10^{4}$ yr ($f_D=10$), respectively. This large discrepancy can be explained once again by noting that these tests represent the two extreme cases with respect to the exponential profile and therefore define the lower and upper limits for the estimate of $\tau_{dep}$.\\
For the exponential-model the best fit $t_{age}$ is $2.1 \times 10^{4}$ yr, while at $\Rd$, $n($\HH$) = 1.7 \times 10^5$ cm$^{-3}$ which corresponds to a $\tau_{dep} = 5.8\times 10^{3}$ yr. As mentioned above, for this model only the adsorption process of CO is included, neglecting desorption at T $\gtrsim 20$ K. Nevertheless, the estimated $t_{age}$ and the corresponding lower limit of $\tau_{dep}$ suggest that the contribution of the adsorption process is negligible at these scales. Furthermore, both ages seem in good agreement with the values of $\tau_{dep}$ provided by the step-function profiles, suggesting again that this is a reasonable approximation to describe the profile of $\chi_{\rm C^{18}O}$.
\end{itemize}

\section{conclusions}\label{conclusions}
\noindent
In order to add a new piece of information to the intriguing puzzle of the depletion phenomenon and its consequences on the chemical evolution in star-forming regions, in the first part of this work we presented and discussed the depletion factor derived for the whole observed structure of G351.77-0.51. In the second part, we reported the results of the estimated depletion radius, as obtained from a simple toy-model described in Sect.~\ref{model}, testing different hypotheses (Sect.~\ref{caveats}).\\
\indent G351.77-0.51, with its proximity and its filamentary structure, seems to be the perfect candidate for this type of study.
We use {\it Herschel} and LABOCA continuum data together with APEX J(2-1) line observations of \co, to derive the $f_D$ map (Fig.~\ref{fig:depletion_map}). 
Along the main body and close to clump-3 the observed depletion factor reaches values as large as $\sim$ 6, whereas in lower density higher temperatures structures $f_D$ is close to 1, as expected.\\
\indent We constructed a simple model that could reproduce the observed $f_D$ in two clumps along the main ridge and along the filament itself.\\
\indent As results of this work, we conclude the following: 

\begin{enumerate}
\item In many regions of the spine and the branches of G351.77-0.51, the depletion factor reaches values larger than $\sim$2.5. This highlights that even in the less prominent structures, as the branches, the depletion of CO can start to occur, altering the chemistry of the inter stellar medium and making it more difficult to study the gas dynamics and to estimate the mass of cold molecular gas (using CO);

\item We find that CO-depletion in high-mass star forming regions affects not only the densest regions of the clumps, but also the filamentary structures that surround them;

\item The model assumed to estimate the size of the depletion radius suggests that it ranges between 0.15 and 0.08 pc, by changing the depletion degree from 10 to the full depletion state. These estimates agree with the full depletion conditions used in the 3D-simulations of \cite{Koertgen17} and \cite{Koertgen18}. This results highlights that such assumptions are not so far from the limits constrained by the observations. We also show that under certain assumptions, it is possible to estimate the size of the depleted region from $f_D$ (see Fig.~\ref{fig:fDRdep_relation});

\item We verified that by changing the shape of the profile assumed to describe the \co/\HH\:variations inside the filament and clumps, the estimates of the size of $\Rd$ do not change more than a factor of 3. 
This difference was interpreted as the final uncertainty associated with the new estimates of $\Rd$ from the physically more realistic case of the exponential profile, where $\Rd \sim 0.04-0.05$ pc;

\item At $R=\Rd$ the model shows a number densities of \HH~between 0.2 and 5.5$\times 10^{5}$ cm$^{-3}$. Following \cite{Caselli99}, we estimated the characteristic depletion time scale for the clump-5, the clump-7 and the filament region included between them (i.e. $\tau_{dep}\sim 5$ and $0.2 \times 10^{4}$ yr, respectively). It is interesting to note that at similar ages, in \cite{Koertgen17, Koertgen18} the simulated deuterium degree also suggest a regime of high depletion on scales that are consistent with the $\Rd$ estimated by our model.
\end{enumerate}

\noindent
Observations at higher resolutions (even in the case of G351.77-0.51) are necessary to put stronger constraints on the models.
Nevertheless, that our results are comparable with the modeling presented by \cite{Koertgen17, Koertgen18} suggests that even if on a large scale it is necessary to include a detailed description of the depletion processes, on the sub-parsec scales we can exploit the conditions of total depletion used in the simulations. This assumption does not seem to strongly alter the expected results but has the enormous advantage of a considerable decrease of computational costs of the 3D-simulations. It would however be necessary in the future to explore the effect of properly modelling the depletion in 3D-simulations, to confirm the findings reported in this work.

\section*{Acknowledgments}
{\rm The authors wish to thank an anonymous referee for their comments and suggestions that have helped clarify many aspects of this work and P.Caselli for fruitful scientific discussions and feedbacks.\\
This paper is based on data acquired with the Atacama Pathfinder EXperiment (APEX). APEX is a collaboration between the Max Planck Institute for Radioastronomy, the European Southern Observatory, and the Onsala Space Observatory. This work was also partly supported by INAF through the grant Fondi mainstream ``{\it Heritage of the current revolution in star formation: the Star-forming filamentary Structures in our Galaxy''}. This research made use of Astropy, a community-developed core Python package for Astronomy (\citealt{Astropy13, Astropy18}; see also \url{http://www.astropy.org}), of NASA’s Astrophysics Data System Bibliographic Services (ADS), of Matplotlib (\citealt{Matplotlib07}) and Pandas (\citealt{Pandas10}). GS acknowledges R. Pascale for feedbacks about Python-improvements. SB acknowledge for funds through BASAL Centro de Astrofisica y Tecnologias Afines (CATA) AFB-17002, Fondecyt Iniciaci\'on (project code 11170268), and PCI Redes Internacionales para Investigadores(as) en Etapa Inicial (project number REDI170093).}

\bibliographystyle{mnras}
\bibliography{mybib_GAL}

\label{lastpage}

\end{document}